\PassOptionsToPackage{table,xcdraw,dvipsnames}{xcolor}
\documentclass[sigconf]{acmart}

\AtBeginDocument{%
  }

\setcopyright{acmlicensed}
\copyrightyear{2018}
\acmYear{2018}
\acmDOI{XXXXXXX.XXXXXXX}

\acmConference[Conference acronym 'XX]{Make sure to enter the correct
  conference title from your rights confirmation email}{June 03--05,
  2018}{Woodstock, NY}

\acmISBN{978-1-4503-XXXX-X/2018/06}

\usepackage{multirow}
\usepackage{framed}
\usepackage{mdframed}
\usepackage{xcolor}
\usepackage{placeins}
\usepackage{tcolorbox}

\begin{document}

\title[``When to Hand Off, When to Work Together'']{``When to Hand Off, When to Work Together'': Understanding Concurrent Human-Agent Interaction in Shared Co-Creative Workspaces.}
\author{Kihoon Son}
\email{kihoon.son@kaist.ac.kr}
\affiliation{%
  \institution{School of Computing, KAIST}
  \city{Daejeon}
  \country{Republic of Korea}
}
\author{Hyewon Lee}
\email{hyewon@purdue.edu}
\affiliation{%
  \institution{Department of Computer Science, Purdue University}
  \city{West Lafayette, Indiana}
  \country{United States}
}
\author{DaEun Choi}
\email{daeun.choi@kaist.ac.kr}
\affiliation{%
  \institution{School of Computing, KAIST}
  \city{Daejeon}
  \country{Republic of Korea}
}
\author{Yoonsu Kim}
\email{yoonsu16@kaist.ac.kr}
\affiliation{%
  \institution{School of Computing, KAIST}
  \city{Daejeon}
  \country{Republic of Korea}
}
\author{Tae Soo Kim}
\email{taesoo.kim@kaist.ac.kr}
\affiliation{%
  \institution{School of Computing, KAIST}
  \city{Daejeon}
  \country{Republic of Korea}
}
\author{Yoonjoo Lee}
\email{lyoonjoo@umich.edu}
\affiliation{%
  \institution{Computer Science and Engineering, University of Michigan}
  \city{Ann Arbor, Michigan}
  \country{United States}
}
\author{John Joon Young Chung}
\email{jchung@midjourney.com}
\affiliation{%
  \institution{Midjourney}
  \city{San Francisco, California}
  \country{United States}
}
\author{HyunJoon Jung}
\email{hjung@mphora.ai}
\affiliation{%
  \institution{Mphora.ai}
  \city{Monte Sereno, California}
  \country{United States}
}
\author{Juho Kim}
\email{juhokim@kaist.ac.kr}
\affiliation{%
  \institution{School of Computing, KAIST}
  \city{Daejeon}
  \country{Republic of Korea}
}
\email{juho@skillbench.com}
\affiliation{%
  \institution{SkillBench}
  \city{Santa Barbara}
  \country{USA}
}

\renewcommand{\shortauthors}{Son et al.}

\newcommand{\sysname}[0]{\textsc{Cleo}}
\definecolor{blockcolor}{HTML}{555555}
\definecolor{blockrule}{gray}{0.6}

\definecolor{handsoff}{HTML}{FBF6B5}
\definecolor{observe}{HTML}{FFCEAD}
\definecolor{concurrent}{HTML}{C2E5FF}
\definecolor{direct}{HTML}{CDF4D3}
\definecolor{terminate}{HTML}{FFC7C2}
\definecolor{border}{HTML}{666666}
\newtcbox{\themetag}[2][]{
  on line,
  boxrule=0.5pt,
  colframe=gray,
  colback=white,
  left=0pt,
  right=0pt,
  top=0pt,
  bottom=0pt,
  fontupper=\footnotesize,
  #1
}
\newcommand{\tagH}{\themetag{handsoff}{H}\hspace{2pt}}
\newcommand{\tagO}{\themetag{observe}{O}\hspace{2pt}}
\newcommand{\tagC}{\themetag{concurrent}{C}\hspace{2pt}}
\newcommand{\tagD}{\themetag{direct}{D}\hspace{2pt}}
\newcommand{\tagT}{\themetag{terminate}{T}\hspace{2pt}}
\newmdenv[
  topline=false,
  bottomline=false,
  rightline=false,
  leftline=true,
  linecolor=blockrule,
  linewidth=1pt,
  innertopmargin=2pt,
  innerbottommargin=2pt,
  innerleftmargin=10pt,
  innerrightmargin=0pt,
  skipabove=10pt,
  skipbelow=10pt,
  backgroundcolor=white,
  font=\small\color{blockcolor}
]{block}

\definecolor{blockcolor}{HTML}{555555}
\definecolor{blockrule}{gray}{0.6}
\newenvironment{prompt}{
  \vspace{5pt}
  \begin{framed}
    \scriptsize\ttfamily\color{blockcolor}
}{
  \end{framed}
}

\begin{abstract}
Shared co-creative workspaces allow users to contribute while agents execute. Yet how users adjust their participation as their work and the agent's execution shape one another remains less understood. We conducted two design probe studies with professional designers. In Study 1 (N=10), participants identified opportunities to contribute by observing the agent's progress, but avoided intervention when the agent misinterpreted their edits. Therefore, we developed \sysname{}, a second probe that tracks user actions and selectively updates its plan. In Study 2 (N=12), we analyzed 255 interaction turns and stimulated recall interviews, identifying five action categories, ten codes, six triggers, and four enabling factors. Participants engaged in concurrent interaction during 33.73\% of turns, expressing newly recognized design ideas or preferences through edits and changing the division of work by completing pending subtasks or independently developing copies of unfinished outputs. We present a decision model and five design implications for concurrent human--agent collaboration.
\end{abstract}

\ccsdesc[500]{Human-centered computing~Interactive systems and tools}

\keywords{Human-Agent Interaction, Interaction Pattern Analysis, Concurrent Interaction, Collaborative Agent, Creative Workflow}

\received{20 February 2007}
\received[revised]{12 March 2009}
\received[accepted]{5 June 2009}

\maketitle

\section{Introduction}

AI agents are increasingly capable of executing complex tasks autonomously~\cite{yang2025magma, yang2025ultra} and support workflows in design~\cite{figmamake, stitch, lovable}, programming~\cite{sapkota2025vibe}, and web navigation~\cite{jiang2025orca}. As agents gain access to shared workspaces through emerging tool-calling infrastructures (e.g., the Model Context Protocol~\cite{anthropic2024mcp}) and their execution becomes visible to users, human--agent collaboration can extend beyond sequential delegation, where users specify goals and review outputs. This opens up opportunities for \textit{concurrent human--agent collaboration}, where users and agents can work simultaneously in a shared workspace. This raises a coordination question: \textit{what interaction patterns emerge during concurrent human--agent co-creation, and what support does this collaboration require?}

Prior work has examined user control during web navigation~\cite{huq2025cowpilot}, editing plans and intermediate research outputs~\cite{feng2024cocoa}, collaboration without prescribed turns in writing and data analysis~\cite{shao2026collaborativegym}, and simultaneous human--AI drawing on a shared canvas~\cite{lawton2023drawing}. However, less is understood about how users adjust their participation when their own work and the agent's execution shape one another. A user's edit can change what the agent should do next, just as the agent's progress can change what the user intends to do. For example, users may edit an element to demonstrate a change for the agent to propagate, complete a pending subtask themselves, or reuse a partial output independently. To inform the design of agents that support concurrent collaboration, we examine when and why users make these contributions, and how they expect the agent to incorporate their edits, adjust the division of work, or preserve their separate work.

\begin{figure}[!htbp]
    \centering
    \includegraphics[width=\linewidth]{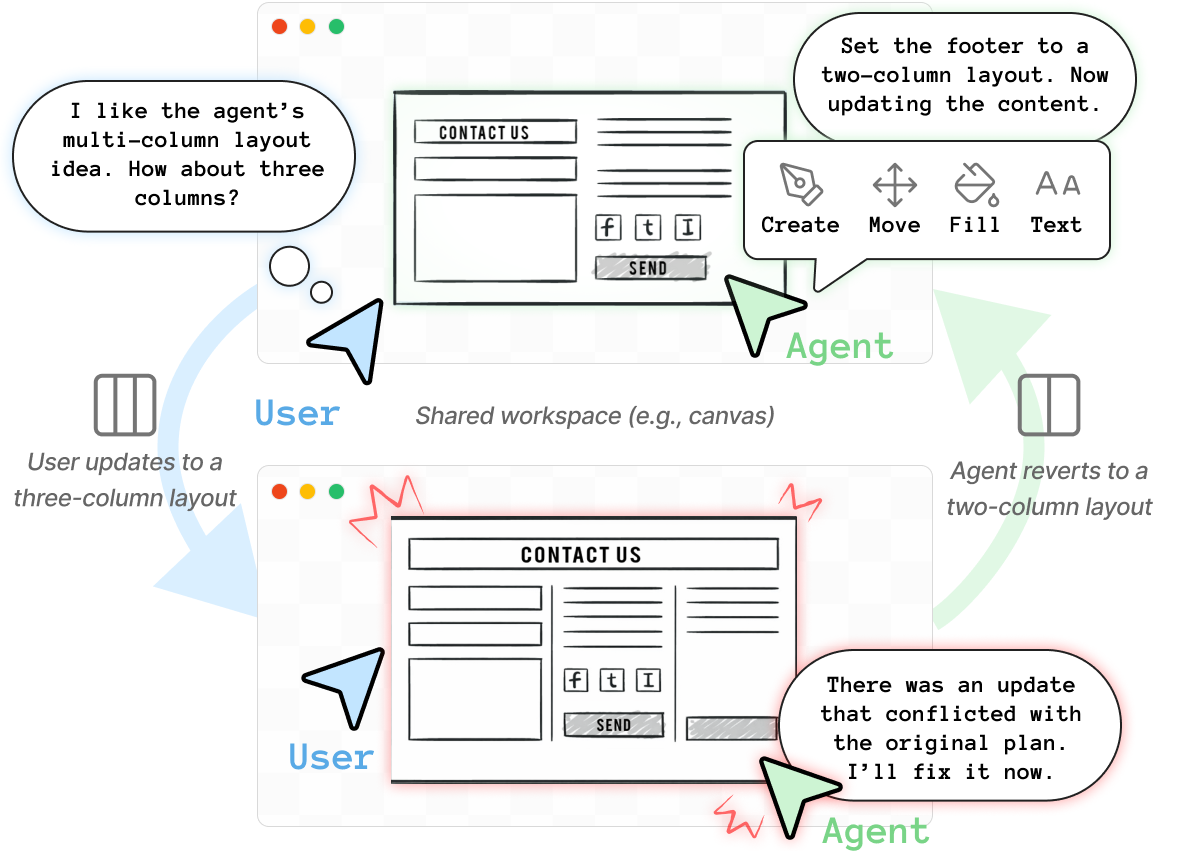}
    \caption{An example of collaborative friction in human-agent collaboration within a shared workspace.}
\Description{Two stacked canvas sketches show conflicting edits to a contact form. At the top, the agent reports setting a two-column layout and updating its content, while the user considers a three-column version. A blue downward arrow shows the user changing the form to three columns in the bottom sketch. The agent then reports that the update conflicts with its original plan and says it will fix it. A green upward arrow shows the agent reverting to the two-column layout. Blue and green cursors identify the user and agent; red spiky marks highlight the conflict around the modified form.}
    \label{fig:problem}
\end{figure}

To explore this, we conducted two design probe studies in a shared co-creative visual workspace, where designers can develop their ideas by directly editing and reusing parts of the agent's output. We first built a process-transparent agent that executes design tasks in Figma through direct tool operations (e.g., creating frames, applying shadows), helping designers track the agent and coordinate accordingly~\cite{gutwin2002descriptive, gutwin1996workspace, robertson2002public}. In a single-session study with 10 professional designers (Study 1), participants used visible progress to evaluate the agent's direction and identify opportunities to contribute. They wanted to reuse intermediate results and demonstrate intent through direct manipulation. However, when users edited the shared canvas while the agent was executing, the agent sometimes misinterpreted or reverted their modifications, creating coordination conflicts and discouraging further intervention (Figure~\ref{fig:problem}). This reluctance to intervene limited us to observe the collaboration strategies participants wanted to pursue. These findings highlighted the need for \textit{collaborative context awareness} on the agent's side, so it could respond to users' contributions in the ways they expected.

Based on these findings, we developed \sysname{} (Collaborative Linked Executive Operator), a second design probe that incorporates collaborative context awareness by tracking user actions within the workspace, interpreting whether these actions should update the agent's task planning or be considered independent work, and adapting subsequent execution accordingly~\cite{horvitz1999principles, pellegrinelli2016human}. We used this probe to investigate how users coordinate their contributions with ongoing agent work and what responses they expect from the agent.

With \sysname{}, we conducted a two-day exploratory study with 12 professional designers (Study 2), analyzing 255 interaction turns and stimulated recall interviews~\cite{dempsey2010stimulated, calderhead1981stimulated}. We identified a taxonomy of \textbf{five action pattern categories and ten codes}, along with \textbf{six triggers} prompting intervention and \textbf{four enabling factors} shaping users' choices among these action patterns. While viewing the agent's intermediate outputs, participants identified design preferences they had not considered beforehand and expressed them through concurrent edits. They also changed the division of work during execution, completing pending subtasks or developing copies of unfinished artifacts independently while the agent continued its assigned work. We synthesized these findings into a decision model with six interaction loops explaining how users move between these action categories during agent execution (Figure~\ref{fig:process_model}).

Drawing from these findings, we derive five design implications for adaptive collaborative agents, such as adapting progress detail to users' needs and updating the agent's remaining plan when users complete pending subtasks. We discuss how concurrent interaction may change as agents become more capable, its potential for workstyle learning, its extension across domains and applications, and the trade-offs of human--agent collaboration. We release the code for \sysname{}, our probe system for concurrent human--agent interaction, to support deeper investigation of how, when, and why users contribute while agents are working. Our contributions include:

\begin{itemize}
    \item A taxonomy of user action patterns, comprising five categories and ten codes characterizing participation in shared workspaces that support concurrent human--agent co-creation.
    \item A decision model with six interaction loops explaining how users navigate between delegation, direction, and concurrent work based on six triggers and four enabling factors.
    \item Design implications for adaptive collaborative agents, covering process visibility adaptation, behavioral signal interpretation, and proactive execution management.
\end{itemize}

\section{Related Work}

\subsection{Participation and Coordination in Shared Workspaces}

People follow their collaborators' progress while considering when to help and when to continue their own work. Coordinating contributions involves managing turns in conversation~\cite{sacks1974simplest}, as well as considering how to share and integrate distributed work and the costs involved in collaborative learning~\cite{kirschner2009cognitive}. In shared workspaces, gestures and changes to artifacts reveal others' activities~\cite{tang1991findings, robertson2002public}, helping collaborators decide whether to build on that work or address another part first. Workspace awareness research describes the information supporting these decisions~\cite{gutwin2002descriptive, gutwin1996workspace, dourish1992awareness}; interfaces that convey gaze and gestures help remote collaborators coordinate their actions~\cite{ishii1993integration, nam2009collaborative}.

These considerations also apply to human--agent collaboration. Users may take on parts of a delegated task or change its direction. Mixed-initiative principles emphasize combining direct manipulation and automation as circumstances demand~\cite{horvitz1999principles}, supported by easy correction~\cite{amershi2019guidelines} and agents' responsiveness to changing needs~\cite{cila2022designing}. Supporting users' workflows therefore requires understanding when they work directly and when they delegate again. We analyze transitions among observation, direct work, and delegation, and the reasons behind them, to inform agent support in shared workspaces.

\subsection{Delegation and Workflow Steering in Human--Agent Collaboration}

Research on human--agent collaboration has explored how users delegate tasks and steer agents' work. LLM tool use and interleaved reasoning and action~\cite{schick2023toolformer, yao2023react} enable multistep delegation in programming~\cite{jimenez2023swe, sapkota2025vibe, claudecode}, web tasks~\cite{deng2023mind2web, kim2024autoIntent, zhou2024proposer}, scientific research and data analysis~\cite{gottweis2025towards, hong2025datainterpreter}, and general digital work~\cite{manus}.

To help users inspect and revise work, researchers have connected AI explanations to the task interface. Talk to the Hand points to relevant on-screen elements with a separate AI cursor and speech bubbles~\cite{prasongpongchai2025talktothe}, while WaitGPT visualizes filtering, aggregation, and intermediate results to help users identify what to change~\cite{xie2024waitgpt}. Tools also incorporate users' judgments into subsequent execution through editing and rewinding~\cite{epperson2025agdebugger} or pausing to clarify ambiguous preferences~\cite{peng2025morae}.

Users can also participate directly in agents' workflows. CowPilot and Orca let users take control of web tasks~\cite{huq2025cowpilot, jiang2025orca}; CowPilot incorporates their action histories when execution resumes. Cocoa supports planning and assigning responsibilities before execution, as well as revising intermediate artifacts and subsequent plans during execution~\cite{feng2024cocoa}. Collaborative Gym examines performance, initiative, collaborative behaviors, and failures in a shared-control environment without prescribed turns~\cite{shao2026collaborativegym}.

However, more opportunities to intervene do not always make collaboration easier. Editing plans and supervising individual steps can correct errors while increasing cognitive burden~\cite{he2025planthenexecute}, and users may want proactive help yet dislike suggestions to change ongoing work~\cite{salikutluk2024situational}. How users intervene also depends on their goals and circumstances. CowCorpus examines intervention motives and timing in web tasks~\cite{huq2026modeling}, while other work links users' goals to participation and explanation needs~\cite{cheng2026mapping}. In ProAIDE, users opened suggestions more often after commits than after rejecting AI edits~\cite{kuo2026developer}. Further, Fairy Cursor explores assistance embedded in ongoing work, highlighting the need to align agent initiative with users' tasks and readiness to engage~\cite{cao2026fairy}.

We build on these studies to examine how users work with intermediate artifacts while an agent continues working. Our work specifically explores when users expect the agent to adopt their edits, continue independently, or leave parts of the task to them.

\subsection{Co-Creative Collaboration through Direct Manipulation}

Creators clarify desired outcomes by developing and revising intermediate results. In design, understanding of a problem evolves alongside possible solutions, shaping the direction of the work~\cite{maher1996modelling}. When using generative and editing tools~\cite{figmamake, stitch, lovart}, creators need to explore beyond initial suggestions given the risks of ideational homogenization and fixation~\cite{anderson2024homogenization, wadinambiarachchi2024theEffects}. Artographer uses generated images as starting points for discovering other artworks, letting users browse an artwork map and collect images of interest~\cite{almeda2025artographer}.

Co-creative systems support participation from selecting and revising outputs to shaping production workflows. Users request reference images while sketching~\cite{karimi2020creative} and compare and combine visual alternatives through dialogue when editing graphics~\cite{zhou2024understanding}. In Figma, they can follow stepwise guidance or automate selected steps~\cite{khurana2025doitfor}. Product designers discuss requirements and proposals with agents~\cite{ding2023designgpt}; animators review scripts and storyboards~\cite{wang2025animagents}; and designers edit and branch work histories to explore alternative directions~\cite{you2025designmanager}. Workflow authors can also configure task order and agent assignments~\cite{hao2025flowforge}.

Working directly with artifacts can make the targets and direction of requested changes concrete. Coauthors review and apply revisions proposed by AI through document comments~\cite{lehmann2026collaborative}, while Reframer users edit drawings as the AI draws~\cite{lawton2023drawing}. AI-Instruments lets users place lenses containing prompts over images to change their style~\cite{riche2025ai}, and DuetUI uses records of users' filter selections to guide subsequent searches and recommendations~\cite{xu2026duetui}. These systems let users steer subsequent AI actions through interaction with artifacts.

In such collaboration, users can adjust what they do and delegate as work progresses, moving from observing an agent to editing part of its work, or demonstrating a direction before delegating the rest. Supporting these transitions requires understanding why users change how they participate and what responses they expect from the agent. We use design work with \sysname{} as a setting to analyze transitions among observing, working directly, giving instructions, and delegating, and the triggers and conditions shaping those choices. We derive design implications for coordinating agent actions with user participation.

\section{Study 1: Initial Technology Probe Study}
To explore how process visibility creates opportunities for co-creative interactions and reveals the coordination challenges it entails in creative workflow, we implemented an initial design probe and conducted an exploratory study with 10 designers.

\subsection{First Probe System}
\subsubsection{Three Basic Components for Collaborative Agent Design}
Existing design agents are largely opaque---they accept instructions and return text reasoning (or high-level sub-goals) and final outputs, giving users little visibility into execution. To make agent's execution observable and collaborative, we grounded the probe's design consideration in three core components drawn from human-human collaboration literature, where these principles have been shown to support coordination, mutual understanding, and fluid collaboration, reasoning that they could equally scaffold more natural human-agent collaboration: (1) \textit{Interaction Modality} for peer-like communication, (2) \textit{Process Transparency} for observable agent work, and (3) \textit{Presence Awareness} for mutual attention visualization.

\textbf{(1) Interaction Modality: Voice Interaction with Flexible Object Referencing.}
The first probe uses voice as the primary interaction channel, enabling natural turn-taking, real-time intention sharing, and a sense of horizontal partnership \cite{hutchins1995cognition, reicherts2022s, bickmore2005establishing}. Users can also reference canvas objects directly during speech, enabling fluid deictic communication (e.g., ``make this larger'' while clicking)~\cite{kim2021winder}.

\textbf{(2) Process Transparency: Sharing Implementation Workflow.}
The first probe system exposes its design process in real-time through step-by-step execution, so users observe not just \textit{what} the agent produces but \textit{how} it progresses, establishing common ground around both goals and process \cite{robertson2002public, cila2022designing, hill1992edit, karaut2002theuseof}.

\textbf{(3) Presence Awareness: Visualizing Attention in the Shared Workspace.}
Following Gutwin's workspace awareness framework \cite{gutwin2002descriptive}, the agent's position reveals its current focus, enabling users to anticipate upcoming actions and coordinate accordingly \cite{dourish1992awareness, fussell2000coordination}.

\subsubsection{System Design}
\hfill
\begin{figure*}[!ht]
    \centering
    \includegraphics[width=0.95\textwidth]{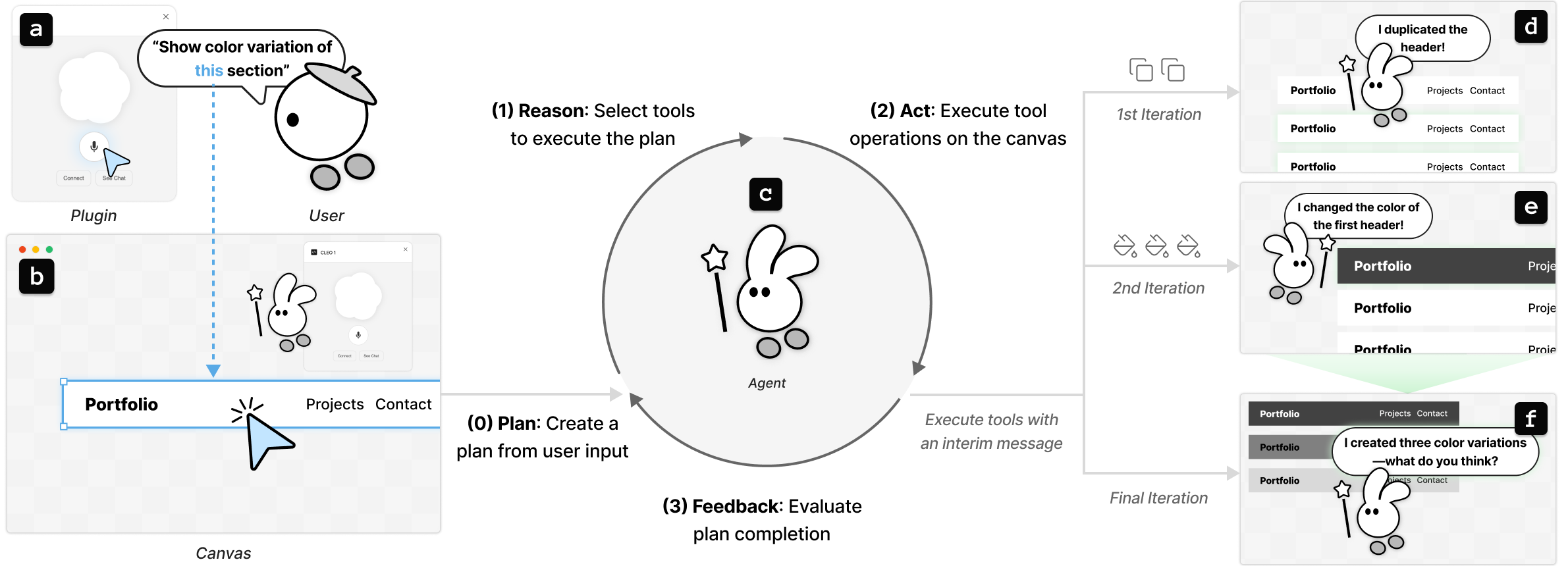}
    \caption{Interaction flow of the first probe system. (a) The user sends a voice message (b) while selecting canvas elements for context. (c) The agent generates an action plan based on the voice transcription with selections and iterates through ReAct cycles. (d-f) Each iteration's progress is reflected immediately on the canvas as tools execute during the act stage.}
\Description{Panels a and b show a user requesting color variations by voice while selecting a portfolio header on the canvas. Panel c shows a plan generated from this input followed by a repeating cycle: reason to select tools, act by executing canvas operations, and obtain feedback on plan completion. Panels d through f show the visible results with an agent character and progress messages. The first iteration duplicates the header, the second changes the color of the first header, and the final iteration presents three color variations.}
    \label{fig:system1}
\end{figure*}
\textit{Interface and Interaction Scenario.}
The first probe is implemented as a Figma plugin, in which users combine voice input with canvas element selection to provide context-based instructions to the agent (Figure \ref{fig:system1}-a-b). The agent responds through iterative tool execution---reasoning, acting, and self-evaluating in cycles---with each step reflected immediately on the canvas via a floating agent character alongside interim messages. (Figure \ref{fig:system1}-c-f).

\textit{Agent Design and Pipeline.}
The first probe agent is built on the ReAct structure \cite{yao2023react}, operating through a three-step iteration cycle: (1) reason, (2) act, and (3) feedback (Figure \ref{fig:system1}-c; Appendix \ref{appendix:system1pipe}). At the reasoning stage, the agent selects from 38 Figma operations (e.g., create rectangle) and generates an interim progress message, which is executed on the canvas alongside the agent character. After each cycle, the summary module generates a text description of what occurred, while the feedback module compares the canvas state before and after execution to evaluate plan completion. If the plan is fulfilled, the agent terminates and delivers a final response; otherwise, it begins a new iteration, up to a maximum of 10 cycles \label{maxturn}. 

We use Claude Sonnet 4.5~\cite{anthropic2025sonnet} as the main reasoning model and Claude Haiku 4.5~\cite{anthropic2025haiku} for all other modules. Full prompt instructions, including Figma design guidelines (adapted from~\cite{jeong2026canvas}) and the complete set of 38 Figma operation tools (See Appendix \ref{appendix:tool_list}).

\subsection{Study Design}
Study 1 investigates two questions: (1) How does process transparency---observing the agent's step-by-step workflow on the canvas---affect users' collaboration experience and enable a new type of collaboration? (2) What design considerations are needed for better human-agent collaboration in co-creative workflow? We observed 10 professional UI/UX designers collaborating with our design probe across different task types to examine how process transparency shapes their understanding and collaboration strategies.

\subsubsection{Participants}
We recruited 10 UI/UX designers (D1-D10, 3 male, 7 female, ages 24-36, mean=28.30) with at least 1 year of professional design experience. We screened for high Figma proficiency (e.g., daily users) and prior experience with diverse design agents (e.g., Figma Make~\cite{figmamake} or Lovart~\cite{lovart}). Participants were compensated 40,000 KRW for approximately 1.5 hours of participation.

\subsubsection{Procedure}
After introducing different types of design agents (e.g., OptiMuse~\cite{zhou2024understanding} and Figma Make~\cite{figmamake}) and a brief tutorial on the probe's three interaction components, participants completed two design tasks of varying scope to observe how collaboration strategies differ across task complexity: (1) a fine-grained task designing a single webpage section (e.g., hero or project gallery) with given materials (15 mins), and (2) a coarse-grained task creating an entire webpage (e.g., startup landing page) across multiple sections (25 mins). We concluded with a 30-minute semi-structured interview exploring how process transparency shaped their collaboration strategies and interaction with the probe.

\subsection{Findings}
\subsubsection{Process Transparency: Real-Time Understanding and Intervention Opportunities}

Observing the agent's step-by-step execution fundamentally transformed how designers experienced collaboration. All designers reported a strong sense of ``working with the agent'' rather than delegating to then reviewing it. Unlike traditional agents, where users assign tasks and wait for final outputs, process transparency enables concurrent work: participants understand what the agent is doing, how it will end up, and plan or redirect their own tasks based on this understanding.

This transparency relieved the ``loss of agency'' problem common in human-agent interaction \cite{schombs2026fromconv}. Participants contrasted our probe with general design agents, where agents take over the workspace, leaving users to wait passively and fearing disruption if they intervene. With those agents, users must read lengthy reasoning texts to understand their plans or reasoning---a cognitively demanding process. In contrast, seeing intermediate behaviors with outputs on the canvas made the agent's intentions intuitively understandable. D4 commented ``I don't need to read a paragraph. I just see it creating a button, and I know what's happening.'' This understanding of how the agent works prevented over-dependence on opaque automation, preserving user agency throughout the process.

Most critically, step-by-step observation enabled early evaluation and proactive intervention. Participants could anticipate the agent's next steps (D1-D10), identify the agent's misinterpretation on the given task (D1, D3, and D5-D7), recognize their own unclear instructions (D1-D2, D4, and D8), and assess the agent's overall workflow direction (D2, D6, and D8-D10)---all before the agent finished. Importantly, this understanding enabled participants to plan their own flexible parallel work: while the agent executed, participants gathered reference images, evaluated overall design directions, or prepared content for subsequent sections, rather than passively waiting. Specifically, D7 stated: ``I saw it working on the hero section with a certain direction, so I started looking for references related to that direction for the next section instead of just watching.'' This predictability helped participants manage workflows more effectively, deciding when to intervene, to prepare subsequent tasks, and to let the agent continue autonomously. Process transparency thus shifted collaboration from reactive correction of final outputs to proactive coordination during execution.

\subsubsection{Emerging Expectations: Flexible Concurrent Collaboration and Context Awareness}
Agent's process transparency fundamentally changed users' expectations: they wanted to engage directly with the agent's work-in-progress through concurrent intervention rather than waiting for completion. For example, participants wanted to appropriate intermediate outputs for their own designs, assist the agent by working on related elements simultaneously, or demonstrate desired changes through direct editing. Most of the participants (D1-D3 and D6-D10) explained that observing the process made them more aware of how they wanted things done, creating natural moments to intervene and collaborate. D3 noted: ``I could see exactly where it needed my input---I wanted to just jump in and show it rather than explain this verbally.''

However, the first probe's inability to distinguish between user and agent actions on the canvas led to conflicts. Because the agent tracked overall canvas changes without separating user modifications, it sometimes reverted or misinterpreted concurrent user work. Participants who experienced this avoided intervening during execution, relegating their collaboration ideas to post-task interviews. Yet even participants who avoided intervention reported that their reluctance to interrupt agents stems from prior experiences with other agent systems (D2, D4-D7, D10)---fearing that mid-execution changes would derail the agent's workflow. Process transparency challenged this assumption; designers recognized opportune moments for intervention but lacked system support.

Participants envisioned real-time interleaved workflows where users could intervene during the agent's ongoing execution: appropriating agent-created elements into their own designs while the agent continued working, handing off their work-in-progress for the agent to build upon mid-execution, or collaboratively editing the same artifact simultaneously. Critically, they expected agents ``to understand why they intervened''---whether corrections indicated misunderstanding, appropriations showed approval, or demonstrations communicated desired style (D1, D3-D5, D7-D8). This requires the agent's collaborative context awareness: distinguishing feedback on the agent's work from independent user work and recognizing when users want the agent to adapt versus continue autonomously. Process transparency thus revealed not just new collaboration opportunities, but also the need for context-aware agents that support --- rather than conflict with --- real-time interleaved human actions.

\section{Study 2: Two-Day Technology Probe Study with Stimulated Recall Interviews}

While Study 1 revealed the coordination problem that arises when agents lack collaborative context awareness, it remained unclear what interaction patterns emerge in the collaboration and what enables users to shift between delegation, direction, and concurrent work. To investigate this, we conducted a two-day technology probe study~\cite{hutchinson2003technologyprobes} with stimulated recall interviews~\cite{dempsey2010stimulated, calderhead1981stimulated} observing experienced creative professionals collaborate with the probe.

\subsection{Second Probe System: \sysname{} \\(Collaborative Linked Executive Operator)}
\subsubsection{Understanding Collaborative Context for Flexible Human-Agent Co-Creation}

Study 1 revealed that process transparency prompts users to interleave their actions with the agent's ongoing execution. The key challenge is collaborative context awareness---agents must distinguish whether concurrent user actions represent feedback requiring adaptation, independent parallel work to be preserved, or co-editing to be incorporated into ongoing execution.

Therefore, we developed \sysname{} (Collaborative Linked Executive Operator), which extends the first probe by incorporating collaborative context awareness as its core principle, enabling the agent to recognize and adapt to users' direct manipulation during execution.

\subsubsection{\sysname{} Design and Improvements from Design Probe 1}
\hfill
\begin{figure*}[h]
    \centering
    \includegraphics[width=\linewidth]{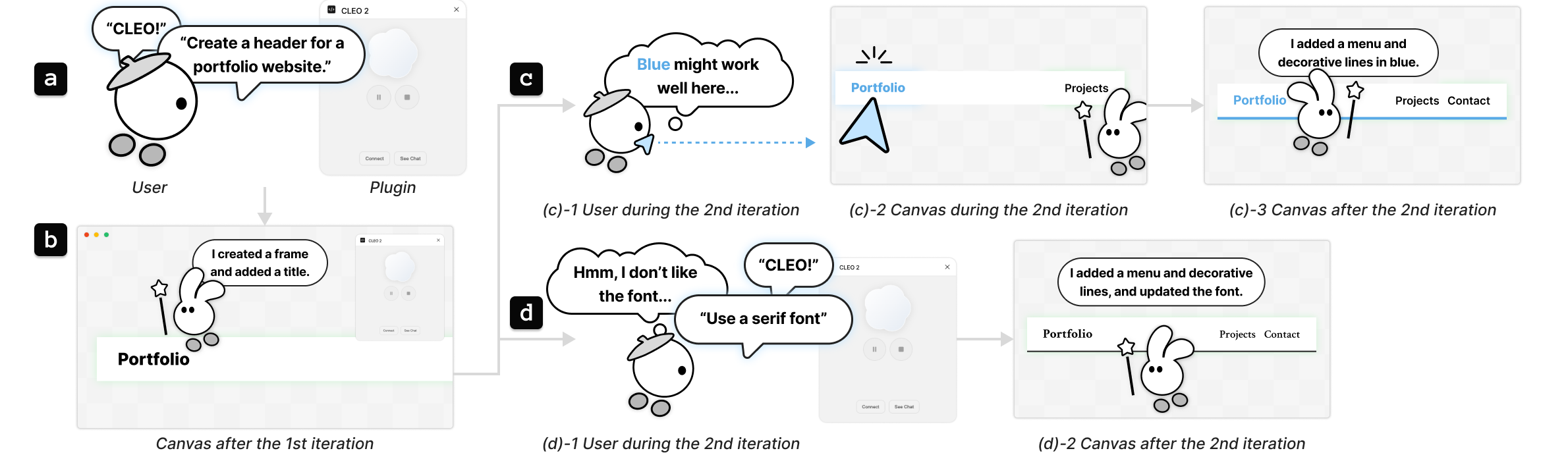}
    \caption{Flexible co-creation scenario with \sysname{}. (a) The user invokes \sysname{} by calling its name. (b) While \sysname{} is acting on the canvas, the user can (c) work concurrently on the same task, or (d) intervene directly with additional instructions at any time.}
\Description{Panel a shows the user calling CLEO and requesting a portfolio header. Panel b shows the first iteration's result: a frame and a Portfolio title. Two alternative branches illustrate the second iteration. In panel c, the user changes the title to blue while the agent works; the agent then adds menu items and decorative lines in blue. In panel d, the user calls CLEO and requests a serif font; the resulting header contains the menu, decorative lines, and updated serif text. User characters and blue cursors mark the user's contributions, while rabbit characters and speech bubbles show the agent's responses.}
    \label{fig:system2}
\end{figure*}

\textit{Interaction Improvements.}
Building on the first probe system, \sysname{} supports three additional interaction types (Figure \ref{fig:system2}). First, users can abort the agent's operation at any time via the termination button on the plugin. Second, users can work concurrently with the agent on the same elements, with the agent tracking and prioritizing user operations to align its plan accordingly (Figure \ref{fig:system2}-c). Third, users can provide additional voice instructions at any time by calling the agent's name while it is working, which the agent incorporates in the next iteration (Figure \ref{fig:system2}-d).

\begin{figure*}[h]
    \centering
    \includegraphics[width=\linewidth]{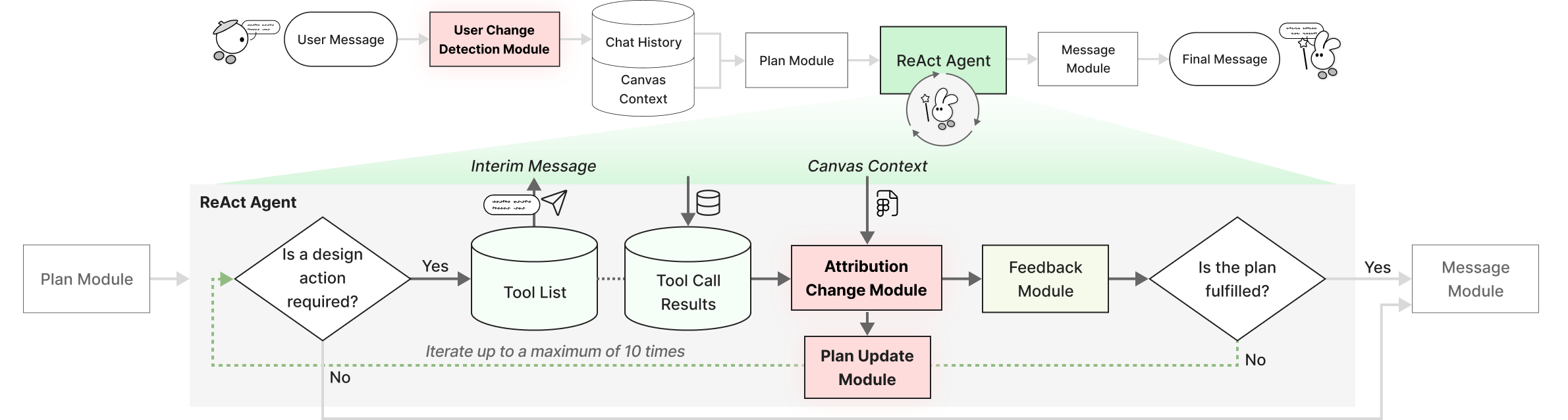}
    \caption{Agent structure of \sysname{}. Three modules have been updated from the first probe pipeline (Appendix \ref{appendix:system1pipe}): User Change Detection Module, Attribution Change Module, and Plan Update Module.}
\Description{The top pipeline runs from User Message through User Change Detection, Chat History and Canvas Context, the Plan Module, the ReAct Agent, and the Message Module to a Final Message. The expanded ReAct flow below first asks whether a design action is required. If not, it goes directly to the Message Module. If yes, tools execute, producing interim messages and tool-call results. These results and canvas context feed the Attribution Change Module, which connects to the Feedback Module and branches to the Plan Update Module. A plan-fulfillment check sends completed work to the Message Module or repeats the loop, up to ten iterations. Pink boxes highlight User Change Detection, Attribution Change, and Plan Update.}
    \label{fig:system2pipe}
\end{figure*}

\textit{Technical Improvements.}
To support these, we replaced the original \textit{Summary Module} with three new modules (Figure~\ref{fig:system2pipe}). The \textit{User Change Detection Module} analyzes user actions that occurred while the agent was inactive, maintaining awareness of the overall workflow. The \textit{Attribution Change Module} distinguishes between agent and user actions during execution, producing separate summaries for each. The \textit{Plan Update Module} then updates the current plan only when user actions conflict with or indicate new directions from the existing plan, before initiating the next iteration cycle.

\begin{figure}[h]
    \centering
    \includegraphics[width=\linewidth]{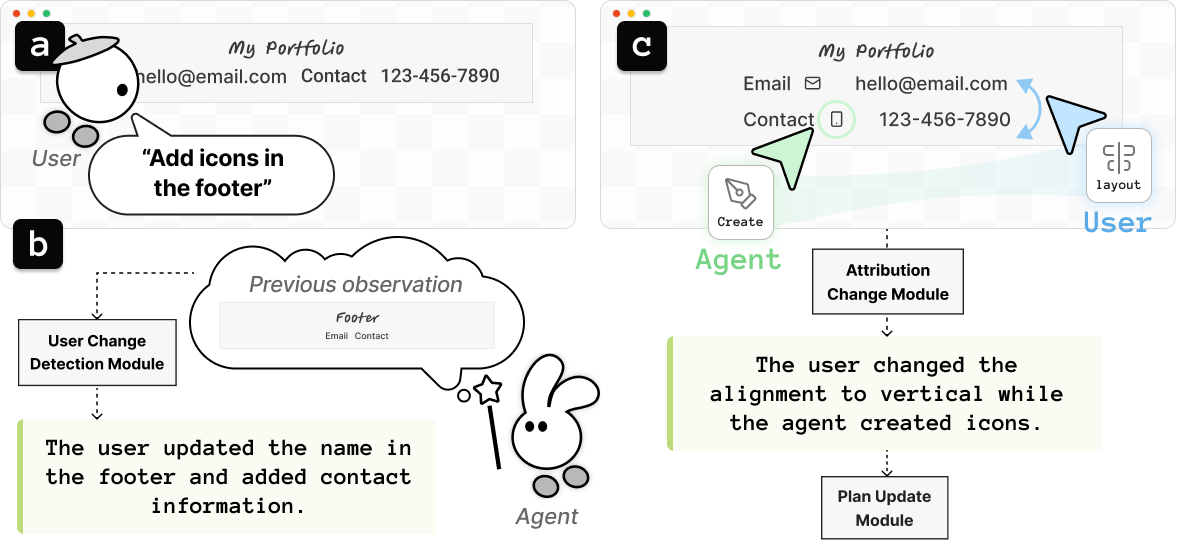}
    \caption{How \sysname{} achieves context awareness. (a) The user invokes the agent. (b) It recalls past observations and detects user updates against the current state. (c) It monitors concurrent user actions while executing its plan, prioritizes interventions, and updates its plan as needed.}
    \Description{The figure demonstrates how the system achieves context awareness through three components. In (a), the user invokes the agent by issuing a voice command to add icons to the footer of a portfolio canvas. In (b), the User Change Detection Module compares the agent's previous observation — a simple footer with email and contact fields — against the current canvas state, detecting that the user has updated the footer content and added contact information. In (c), while the agent is actively creating icons on the canvas, the Attribution Change Module detects that the user concurrently changed the layout alignment to vertical. This intervention triggers the Plan Update Module, allowing the agent to revise its plan accordingly.}
    \label{fig:context}
\end{figure}

\subsection{Study Design}
We conducted a two-day exploratory study with \sysname{} to understand what interaction patterns emerge when users collaborate with a context-aware agent, and what enables or triggers different collaboration modes. By combining interaction logs with stimulated recall interviews, we aimed to capture not just what users did, but why they made those choices in context.

On the first day, participants performed two design tasks (closed- and open-ended) in Figma, with interactions logged and screen-recorded. After the session, three authors visually represented the interaction logs by segmenting the entire task process into agent-activated and agent-deactivated periods, and labeling each segment with timestamped user and agent actions to create an easily reviewable format for the stimulated recall interview ( Appendix \ref{appendix:vis_method}-c-d). We gave a one-day gap between the task session and recall interview to avoid participant fatigue and for visualizing logs. On the second day, stimulated recall interviews were conducted where participants reviewed these visualized interaction logs.

\subsubsection{Participants}
We recruited 12 participants (4 males and 8 females; M\_age = 29.8, SD\_age = 3.7) who met two key criteria: (1) at least two years of professional design experience, and (2) active use of VLMs (e.g., ChatGPT~\cite{openai2025chatgpt} or Gemini~\cite{google2025gemini}) or design agents (e.g., Figma Make~\cite{figmamake}, Stitch~\cite{stitch}, or Lovable~\cite{lovable}) in their creative workflow. We recruited expert designers who have an average of 4.2 years of professional experience (SD = 1.7) in the design field. Participants were compensated 120,000 KRW for approximately 3 hours of participation.

\subsubsection{Procedure}
\hfill

\textit{First day.} After an introduction, participants familiarized themselves with \sysname{} through a warm-up task (30 mins), followed by two main tasks: (1) a closed-ended task (30 mins) where participants designed an empty section of a web portfolio to match existing sections given design requirements and source materials, and (2) an open-ended task (40 mins) where participants selected one of three startup concepts and freely designed a webpage from scratch.

After the tasks, three authors processed the collected interaction logs by marking timestamps of user requests and agent task completion. Using these timestamps, each participant's entire task process was segmented into agent-activated and agent-deactivated periods, then grouped by each user request. Segmented video clips were also prepared alongside these logs.

After these tasks concluded, three authors prepared visualizations of the interaction dynamics to facilitate the stimulated recall interview on the second day. The authors first marked the moments when participants sent requests to the agent and when the agent completed execution for each request (Appendix \ref{appendix:vis_method}-a). Using these two time points to segment the entire task workflow, the authors visualized the data (Appendix \ref{appendix:vis_method}-b-d), distinguishing between periods when the agent was actively working and when it was not. Alongside these visualizations, we prepared segmented video clips of the actual task recordings.

\textit{Second day.} Participants reviewed their visualized interaction logs and corresponding video clips in a stimulated recall interview lasting up to one hour. For each segment, we asked what they were doing in the clip, their reasons or purposes for acting that way, and the collaborative situations in which they considered such behavior necessary.

A 20-minute wrap-up interview followed, covering participants' overall collaboration experiences, changes in how they worked with the agent between the two tasks, and comparisons with other design agents (Appendix~\ref{appendix:study2_interview_protocols}).

\subsubsection{Analysis Methods}
\hfill

\textit{Qualitative coding of the user's action patterns.}
From the study, we collected 255 turn-level data from 12 participants across two tasks, where a turn means a complete agent execution in response to a user request comprising up to 10 iterations (Section \ref{maxturn}), with an average of 21.25 turns per participant (min=12, max=33). Three authors conducted open coding~\cite{corbin1990grounded, khandkar2009open} on what actions users performed while the agent was working, separating multiple actions within a single turn for individual coding. Two authors iteratively reviewed the data to consolidate overlapping actions into broader categories, after which a third author joined to refine and finalize until consensus was reached. As a result, we defined 5 categories and 10 codes (Table \ref{tab:action_pattern}).

\textit{Distribution analysis of user action patterns.}
After coding, we annotated each of the 255 turns with the 5 categories, where each turn could be mapped to 1 to 5 categories (Appendix \ref{appendix:distribution}), and calculated each category's proportion across all 255 turns.

\textit{Qualitative coding of triggers and enabling factors of user actions.}
Next, we conducted qualitative coding of the stimulated recall interview data to understand when and why users chose different action categories. We coded the task characteristics and situational conditions under which each action category occurred. To develop the enabling factor categories, we examined participants' interview accounts alongside the interaction logs for the corresponding turns. Through this analysis, we identified two key pattern types: \textbf{triggers} that prompted transitions from observation to intervention, and \textbf{enabling factors}---situational conditions including users' mental models, task priorities, and intervention preferences---that determined which specific action users adopted. The coding followed the iterative procedure described above, with a fourth author who had not been involved in conducting the study sessions joining to provide an independent perspective and reduce potential bias.

\textit{Iterative decision modeling of human-agent co-creative collaboration.}
Building on the identified action patterns, triggers, and enabling factors, we constructed an initial decision model describing how user-agent collaboration unfolds from task initiation to completion, representing transitions between different user actions and the conditions under which these transitions occur. Drawing on constant comparison~\cite{corbin1990grounded, glaser2017discovery}, we iteratively compared the developing model with the action sequences observed across all 255 turns. 

When an observed sequence was not adequately explained, we re-examined its associated triggers and enabling factors and the model's transition logic, revising the model to account for the case. This process involved five authors, including an additional author who had not been involved in conducting the study sessions and contributed an independent perspective. Through repeated comparison and discussion, we refined the model to account for the interaction patterns and variations observed in this dataset.

\subsection{Results}
We present five categories of user action patterns and their distribution across 255 turns, followed by six triggers and four enabling factors that explain when and why users chose each pattern, and synthesize these findings into a decision model of human-agent co-creative interaction.

\subsubsection{User Action Patterns During Agent Execution}
\begin{table*}[]
\renewcommand{\arraystretch}{1.2}
\resizebox{0.9\textwidth}{!}{%
\begin{tabular}{llll}
\hline
\rowcolor[HTML]{EFEFEF} 
\textbf{Category} & \textbf{Code} & \textbf{Definition} & \textbf{\% of Turns} \\ \hline
\textbf{(H) Hands-off} & Full delegation & Completely disengaging from the agent's work to focus on other tasks. & 69.41\% \\ \hline
\textbf{(O) Observational} & Observational monitoring & Watching the agent's execution without intervening. & 70.20\% \\ \hline
\textbf{(T) Terminating} & Execution termination & Stopping the agent's work before completion. & 7.06\% \\ \hline
\multirow{2}{*}{\textbf{(D) Directive}} & Instruction-based steering & Providing verbal guidance to adjust the agent's approach. & \multirow{2}{*}{26.67\%} \\
 & Switching tasks & Assigning a new task unrelated to the current execution. &  \\ \hline
\multirow{5}{*}{\textbf{(C) Concurrent}} & Intermediate result appropriation & Copying and using the agent's partial outputs while it continues working. & \multirow{5}{*}{33.73\%} \\
 & Artifact takeover & Duplicating the agent's work-in-progress to edit independently elsewhere. &  \\
 & In-situ co-editing & Working simultaneously with the agent on the same subtask and artifact. &  \\
 & Opportunistic takeover & Completing one of the agent's pending subtasks while it handles another. &  \\
 & Demonstration-based steering & Showing desired changes through direct editing rather than instructions. &  \\ \hline
\end{tabular}%
}
\caption{Five categories and ten codes identified through qualitative analysis of user action patterns. See Figure~\ref{fig:action_pattern} in the appendix for illustrations of each interaction.}
\Description{The table defines ten user action codes in five categories and reports the percentage of 255 turns containing each category. Hands-off (H), 69.41\%: full delegation, in which users disengage from the agent's work to focus on other tasks. Observational (O), 70.20\%: monitoring execution without intervening. Terminating (T), 7.06\%: stopping execution before completion. Directive (D), 26.67\%: instruction-based steering or switching the agent to an unrelated task. Concurrent (C), 33.73\%: intermediate result appropriation, copying partial outputs for use elsewhere; artifact takeover, duplicating work-in-progress for independent editing; in-situ co-editing, editing the same subtask and artifact alongside the agent; opportunistic takeover, completing a pending subtask while the agent handles another; and demonstration-based steering, showing desired changes through direct editing. Multiple categories can occur within a turn, so their percentages sum to more than 100\%.}
\label{tab:action_pattern}
\vspace{-16pt}
\end{table*}

Through the first qualitative coding process, we identified 5 categories and 10 action patterns users took while the agent was working (Table \ref{tab:action_pattern}). Hereafter, each category is denoted by its initial: \textbf{\tagH Hands-off}, \textbf{\tagO Observational}, \textbf{\tagT Terminating}, \textbf{\tagD Directive}, and \textbf{\tagC Concurrent}. These patterns can be characterized along four key dimensions: intervention modality (whether users intervene through verbal instructions or direct actions on the agent's workflow), artifact coupling (whether the user and agent work on separate or overlapping artifacts), synchronicity (whether interactions occur synchronously or asynchronously with agent execution), and task autonomy (who has primary control over the task execution).

Notably, the \textbf{\tagC} category represents a novel set of action patterns enabled by the agent's collaborative context-awareness (Appendix \ref{appendix:action_pattern}). Unlike traditional design agents that block user intervention during execution \cite{figmamake, lovable} or coding agents where direct interventions are not reflected in the agent's work \cite{claudecode, cursorcode}, \sysname{} allows users to directly engage with the agent's ongoing work without disrupting its workflow. This enables more precise communication of nuanced and situational intent---particularly for aesthetic decisions that are difficult to articulate verbally---as observing the agent's step-by-step execution often surfaced latent preferences users had not previously considered. Many participants noted that direct manipulation was easier than verbal explanation for such fine-grained adjustments, prompting them to show the agent what they meant directly (P1-P3, P5, P7-P8, and P10).

Within the \textbf{\tagC} category, \textit{In-situ co-editing} and \textit{Opportunistic takeover} both involve synchronous direct manipulation on overlapping artifacts, yet differ in task autonomy: the former creates shared control over the agent's task (Agent → Shared) while the latter maintains the agent's autonomy over its assigned subtask as the user completes a different pending subtask. Similarly, \textit{Artifact takeover} and \textit{Intermediate result appropriation} both shift to shared control but differ in scope: the former involves duplicating the agent's entire work-in-progress to edit independently, while the latter involves copying only partial outputs.

The \textbf{\tagD} category distinguishes between \textit{Instruction-based steering}, where users provide verbal guidance while maintaining the agent's autonomy, and \textit{Switching tasks}, where users assign a new task while the agent is still executing the previous one.

\textit{Execution termination} of \textbf{\tagT} category shifts task control entirely to the user (Agent → User) by stopping the agent's work completely.
 
Turns ranged from containing only one category to including all five categories (Appendix \ref{appendix:distribution}). Among 255 turns, the most prevalent turn type was immediate hands-off (turns only include \textbf{\tagH}), accounting for 29.8\% of all turns, followed by observation-only turns (turns only include \textbf{\tagO}) at 14.12\%. When we calculated how each action category appeared across 255 turns, \textbf{\tagH} appeared in 69.41\% of turns, \textbf{\tagO} in 70.2\%, \textbf{\tagC} in 33.73\%, \textbf{\tagD} in 26.67\%, and \textbf{\tagT} in 7.06\% (Table \ref{tab:action_pattern}'s \% of Turns).

\begin{table}[!htbp]
\centering
\renewcommand{\arraystretch}{1.0}

\resizebox{\linewidth}{!}{%
\begin{tabular}{lllll}
\hline
\rowcolor[HTML]{EFEFEF} 
Trigger & \textbf{(C)} & \textbf{(D)} & \textbf{(T)} & \textbf{\% of Turns} \\ \hline

\textbf{Idea Spark from Agent's Work-in-Progress} 
& $\checkmark$ & & & 4.31\% \\ \hline

\textbf{Need for Early Outcome Visibility} 
& $\checkmark$ & & & 14.51\% \\ \hline

\textbf{Readiness for Fine-grained Detailing} 
& $\checkmark$ & & & 9.80\% \\ \hline

\textbf{Misaligned Task Interpretation} 
& $\checkmark$ & $\checkmark$ & $\checkmark$ & 10.59\% \\ \hline

\textbf{Execution Quality Drop} 
& $\checkmark$ & $\checkmark$ & $\checkmark$ & 15.29\% \\ \hline

\textbf{Emerging New Task for Agent} 
& & $\checkmark$ & $\checkmark$ & 15.29\% \\ \hline

\end{tabular}%
}

\captionsetup{justification=centering}
\caption{Six triggers that prompt user actions across the
\textbf{\tagC}, \textbf{\tagD}, and \textbf{\tagT} user action categories.}
\Description{Rows list six triggers, with checkmarks indicating the action categories associated with each and a percentage of all 255 turns in which the trigger appeared at least once. C means Concurrent, D Directive, and T Terminating. Idea Spark from Agent's Work-in-Progress: C, 4.31\%. Need for Early Outcome Visibility: C, 14.51\%. Readiness for Fine-grained Detailing: C, 9.80\%. Misaligned Task Interpretation: C, D, and T, 10.59\%. Execution Quality Drop: C, D, and T, 15.29\%. Emerging New Task for Agent: D and T, 15.29\%.}
\label{tab:triggers}
\end{table}

\subsubsection{Triggers and Enabling Factors} Through the second qualitative coding process, we identified six triggers that lead users to perform actions in the \textbf{\tagC}, \textbf{\tagD}, and \textbf{\tagT} categories (Table \ref{tab:triggers}), along with four enabling factor categories that determine which action category users choose when these triggers occur (Table \ref{tab:factors}).

\textbf{\tagO} occurred in all turns except those containing only \textbf{\tagH}, consistently preceding \textbf{\tagC}, \textbf{\tagD}, and \textbf{\tagT} action categories. However, \textbf{\tagO} did not always precede \textbf{\tagH}---some turns showed users initially working hands-off \textbf{\tagH}, then later shifting to \textbf{\tagO}. The \textbf{\tagC}, \textbf{\tagD}, and \textbf{\tagT} categories were triggered during this observation process. Thus, we define triggers as the reasons that lead users to perform one of these three action categories during the observation process.

\begin{table*}[t]
\renewcommand{\arraystretch}{1.2}
\resizebox{\textwidth}{!}{%
\begin{tabular}{llccccc}
\hline
\rowcolor[HTML]{EFEFEF} 
\cellcolor[HTML]{EFEFEF}\textbf{Enabling Factor (Category)} & \textbf{Value (Code)} & \textbf{(C)} & \textbf{(D)} & \textbf{(T)} & \textbf{(O)} & \textbf{(H)} \\ \hline
 & Has mental model & $\checkmark$ & $\checkmark$ & $\checkmark$ & $\checkmark$ & $\checkmark$ \\
\multirow{-2}{*}{\textbf{(a) Mental Model of Agent's Task Capability}} & No mental model &  &  &  & $\checkmark$ &  \\ \hline
 & User task significantly more important &  &  &  &  & $\checkmark$ \\
 & User task moderately more important / Similar importance &  & $\checkmark$ & $\checkmark$ & $\checkmark$ &  \\
\multirow{-3}{*}{\textbf{(b)   Task Importance: User vs. Agent}} & No user task & $\checkmark$ &  & $\checkmark$ & $\checkmark$ &  \\ \hline
 & Verbal explanation easier &  & $\checkmark$ &  & \multicolumn{2}{c}{} \\
 & Direct manipulation easier & $\checkmark$ &  &  & \multicolumn{2}{c}{} \\
\multirow{-3}{*}{\textbf{(c) User's Preferred Intervention Modality}} & Uncertain how to intervene &  &  & $\checkmark$ & \multicolumn{2}{c}{\multirow{-3}{*}{No intervention}} \\ \hline
 & Task will succeed with direct collaboration & $\checkmark$ &  &  & \multicolumn{2}{l}{} \\
 & Agent will understand verbal guidance &  & $\checkmark$ &  & \multicolumn{2}{l}{} \\
\multirow{-3}{*}{\textbf{(d) User's Expectation of Agent's Response to Intervention}} & Agent is incapable of this task &  &  & $\checkmark$ & \multicolumn{2}{l}{\multirow{-3}{*}{No intervention}} \\ \hline
\end{tabular}%
}
\caption{Four categories and eleven codes of enabling factors across action categories.}
\Description{The table groups eleven enabling-factor codes into four categories. Checkmarks associate each code with Concurrent (C), Directive (D), Terminating (T), Observational (O), or Hands-off (H) actions. Mental model of the agent's task capability: having a mental model is associated with all five action categories; having no mental model is associated with O. Importance of the user's task relative to the agent's task: a significantly more important user task is associated with H; a moderately more important or similarly important user task with D, T, and O; and no user task with C, T, and O. Preferred intervention modality: verbal explanation being easier is associated with D, direct manipulation being easier with C, and uncertainty about how to intervene with T. Expected agent response: expecting success with direct collaboration is associated with C, expecting the agent to understand verbal guidance with D, and believing the agent is incapable of the task with T. For the modality and response categories, O and H are marked as no intervention.}
\vspace{-16pt}
\label{tab:factors}
\end{table*}

A total of six triggers were identified (Table \ref{tab:triggers}). Three triggers exclusively led to Concurrent interactions. \textit{Idea Spark from Agent's Work-in-Progress} occurred when observing the agent's intermediate outputs sparked new creative ideas that the user had not previously considered. \textit{Need for Early Outcome Visibility} arose when users wanted to see or utilize the agent's results sooner to plan subsequent tasks. \textit{Readiness for Fine-grained Detailing} emerged when users recognized that the current work stage required detailed refinement aligned with their specific vision.

Two triggers could lead to any of the three action categories (\textbf{\tagC}, \textbf{\tagD}, or \textbf{\tagT}). \textit{Misaligned Task Interpretation} happened when the agent pursued a valid but unintended interpretation of the task, executing in a direction the user did not intend. \textit{Execution Quality Drop} was triggered when the agent's performance fell below the user's acceptable threshold.

Finally, \textit{Emerging New Task for Agent} exclusively led to \textbf{\tagD} or \textbf{\tagT}, occurring when users conceived a new task to assign to the agent while observing the agent's work. Since a given trigger category could appear multiple times within a single turn, we counted the number of turns in which each trigger category appeared at least once. Table \ref{tab:triggers} shows this as \% of Turns across all 255 turns.

We identified four enabling factor categories and eleven codes (Table \ref{tab:factors}). The first factor (Table \ref{tab:factors}-a) concerns whether the user has a mental model of the agent's capability for the assigned task. All action categories except \textbf{\tagO} were performed when a mental model existed, but \textbf{\tagO} also emerged when no mental model was present. For instance, many participants closely monitored the agent's work from start to finish when they lacked a mental model of its capabilities, and even when triggers occurred, they did not proceed to \textbf{\tagT}, \textbf{\tagC}, or \textbf{\tagD} actions in order to first observe how well the agent could perform the given task.

The second factor (Table \ref{tab:factors}-b) concerns the user's perceived importance of their current work relative to the agent's task. When the user's work requires full concentration, they move to \textbf{\tagH}; when importance is similar or the user has no work to attend to, they remain available to observe and intervene, with \textbf{\tagC} occurring when the user has no work and \textbf{\tagD} when their work is of similar or slightly greater importance.

The third factor (Table \ref{tab:factors}-c) concerns the user's preferred intervention modality. When verbal explanation is easier, users choose \textbf{\tagD} actions (e.g., ``Please modify this design based on this reference''). When direct manipulation is easier, users choose \textbf{\tagC} actions, as observed in participants who preferred to directly adjust fine-grained details such as alignment spacing rather than describe them verbally (P1, P3, and P9). When users are uncertain about how to intervene, they choose to terminate the agent's workflow \textbf{\tagT}.

The final factor (Table \ref{tab:factors}-d) concerns the user's expectation of which intervention approach will enable the agent to produce a successful outcome. When users believe direct manipulation alongside the agent will lead to successful completion, they choose \textbf{\tagC}; when they expect verbal feedback alone will be sufficient, they choose \textbf{\tagD}; and when they anticipate the agent cannot perform the task regardless of intervention, they choose \textbf{\tagT}.

\subsubsection{Decision model of Human-Agent Collaboration}
\begin{figure*}[!ht]
  \centering
  \includegraphics[width=\textwidth]{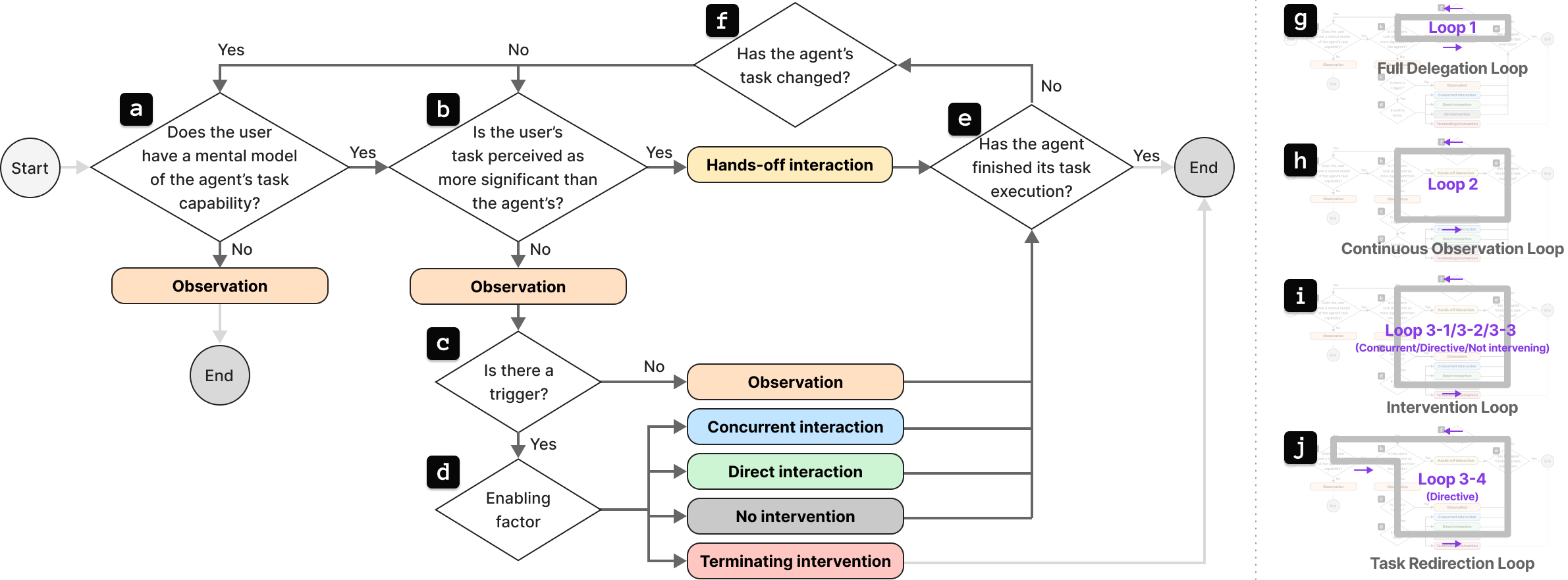}
  \caption{Decision model of human-agent co-creative collaboration}
\Description{A decision flowchart on the left is paired with four loop insets on the right. Decision a asks whether the user has a mental model of the agent's task capability. No leads to observation and an end node; yes proceeds to decision b, comparing task importance to select hands-off interaction or observation. During observation, decision c checks for a trigger. Without one, observation continues; with one, enabling factors at d lead to Concurrent interaction, a green box labeled Direct interaction, No intervention, or Terminating intervention. Decision e checks completion: completed work ends, while unfinished work goes to the task-change check at f. A changed task returns to the mental-model check; an unchanged task returns to the importance check. Termination also ends the flow. Insets g through j highlight six loops: full delegation, continuous observation, three intervention variants for concurrent, directive, or no intervention, and task redirection.}
  \label{fig:process_model}
\end{figure*}

Based on the identified action patterns, triggers, and enabling factors, we synthesized our findings into a \textbf{Decision model} that characterizes how users navigate their interactions with the agent during task execution within a turn (Figure \ref{fig:process_model}). This model represents the implicit decision-making dynamics underlying user behavior, not a rigid, step-by-step conscious flowchart.

The process begins when the agent starts executing a task. The first decision point (Figure \ref{fig:process_model}-a) evaluates whether the user has a mental model of the agent's task capability. If the user lacks this, often occurring when they are new to working with the agent or encountering an unfamiliar task type, they enter \textbf{\tagO} action to build understanding of the agent's capabilities. In this case, observation may continue until the agent completes its execution, allowing the user to develop their mental model for future interactions.

\textit{\textbf{Full delegation Loop.}} If the user already possesses a mental model of the agent's capability for the current task, the process advances to decision point (Figure \ref{fig:process_model}-b), which evaluates whether the user's task is perceived as more significant than the agent's. If so, they move to \textbf{\tagH}; otherwise, they proceed to \textbf{\tagO}.

After entering \textbf{\tagH}, if the agent continues working on its task, the user remains engaged in their own work, creating a loop through two decision points: (1) ``Has the agent finished its task execution?'' (Figure \ref{fig:process_model}-e), and (2) ``Has the agent's task changed?'' (Figure \ref{fig:process_model}-f). If both return ``NO,'' the loop continues back to decision point (b).

This ``Full Delegation Loop'' (Figure \ref{fig:process_model}-g) reveals that hands-off delegation is not a fixed state but a dynamic one. As users make progress on their own work, its relative importance diminishes, causing them to exit \textbf{\tagH} and transition to \textbf{\tagO}, enabling truly parallel work where users fluidly shift between their own tasks and monitoring the agent.

\textit{\textbf{Continuous Observation Loop.}}
During observation, if no trigger occurs, the user continues \textbf{\tagO} through decision points (Figure \ref{fig:process_model}-e and f), forming the ``Continuous Observation Loop'' (Figure \ref{fig:process_model}-h). Participants used continued observation to coordinate their own work with the agent's progress. For instance, P7 and P8 demonstrated strategic monitoring behavior where they timed their observations to coordinate the completion of their own work with the agent's output, planning to combine both results into a unified deliverable. This reveals that observation is an active coordination strategy for managing parallel workflows, not merely passive waiting.

\textit{\textbf{Intervention Loop.}}
When a trigger occurs during observation, the user proceeds to the enabling factors decision, moving toward either \textbf{\tagC}, \textbf{\tagD}, or \textbf{\tagT} actions, or toward ``No Intervention'' when the user has their own work to do and feels that correcting the agent is burdensome (P2, P4, and P9-P10). Aside from \textbf{\tagT}, the other cases form three types of ``Intervention Loop'' (Figure \ref{fig:process_model}-i), where the user's perception of task priorities can change again while performing \textbf{\tagC} or \textbf{\tagD} actions. Specifically, users move to \textbf{\tagC} when they have no other work and believe direct manipulation will lead to more successful task completion, and to \textbf{\tagD} when they have their own work but believe verbal guidance will be sufficient.

Additionally, when a user assigns a new task during \textbf{\tagD} interaction, they move to ``YES'' at ``Has the agent's task changed?'', leading to the ``Task Redirection Loop'' (Figure \ref{fig:process_model}-j).

\section{Discussion}
\subsection{Design Implications for Collaborative Agents in Co-Creative Work}
\begin{figure*}[!ht]
    \centering
    \includegraphics[width=0.9\textwidth]{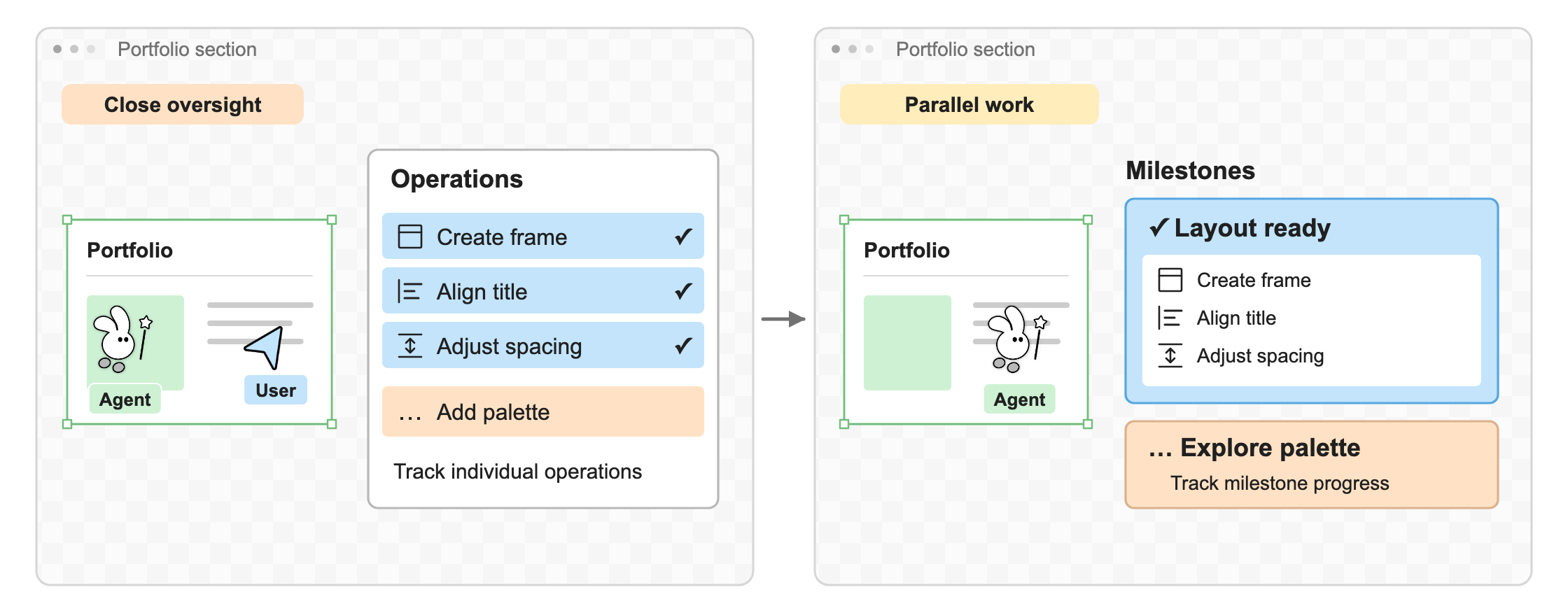}
    \caption{Proposed adaptive process visibility (DI1). The same ongoing execution appears as individual operations for close oversight (left) or grouped milestones for parallel work (right). Both views track progress without requiring step-by-step approval.}
    \Description{Two views of the same portfolio task contrast levels of process detail. The left view lists Create frame, Align title, and Adjust spacing as separate completed operations, followed by Add palette. The right view nests the same three operations inside one Layout ready milestone, followed by Explore palette. Matching blue backgrounds and repeated operation icons show the grouping; peach backgrounds connect the palette work across the views. The labels read Track individual operations and Track milestone progress. Checkmarks indicate completed work, not requests for user approval. Named user cursors and rabbit agent characters identify the collaborators.}
    \label{fig:di1-adaptive-visibility}
\end{figure*}

\textbf{DI1. Adapt process visibility granularity to user needs and task stage for personalized synchronization.}
Agents should structure execution as observable milestones that users can use as synchronization points for timing interventions, delegation, and parallel work. Observational monitoring appeared in 70.2\% of turns, with P7 and P8 actively timing their own work around the agent's subtask completions, directly enabling triggers like ``Idea Spark from Agent's Work-in-Progress'' (Table \ref{tab:triggers}) and reducing clarification costs by catching misunderstandings mid-execution (P1, P4, and P8). However, rather than exposing progress uniformly, agents should adapt granularity to user needs and task stage: finer-grained operations for high-stakes tasks requiring close oversight (P2, P5, and P9), and larger subplan completions for efficiency-focused stages (P6-P8, and P10). This adaptive granularity~\cite{suh2024dynamic} lets users determine their own synchronization rhythm rather than being locked into a single representation pace. Figure~\ref{fig:di1-adaptive-visibility} illustrates this proposed adaptation between operation-level detail and milestone-level summaries.

\begin{figure*}[!ht]
    \centering
    \includegraphics[width=0.9\linewidth]{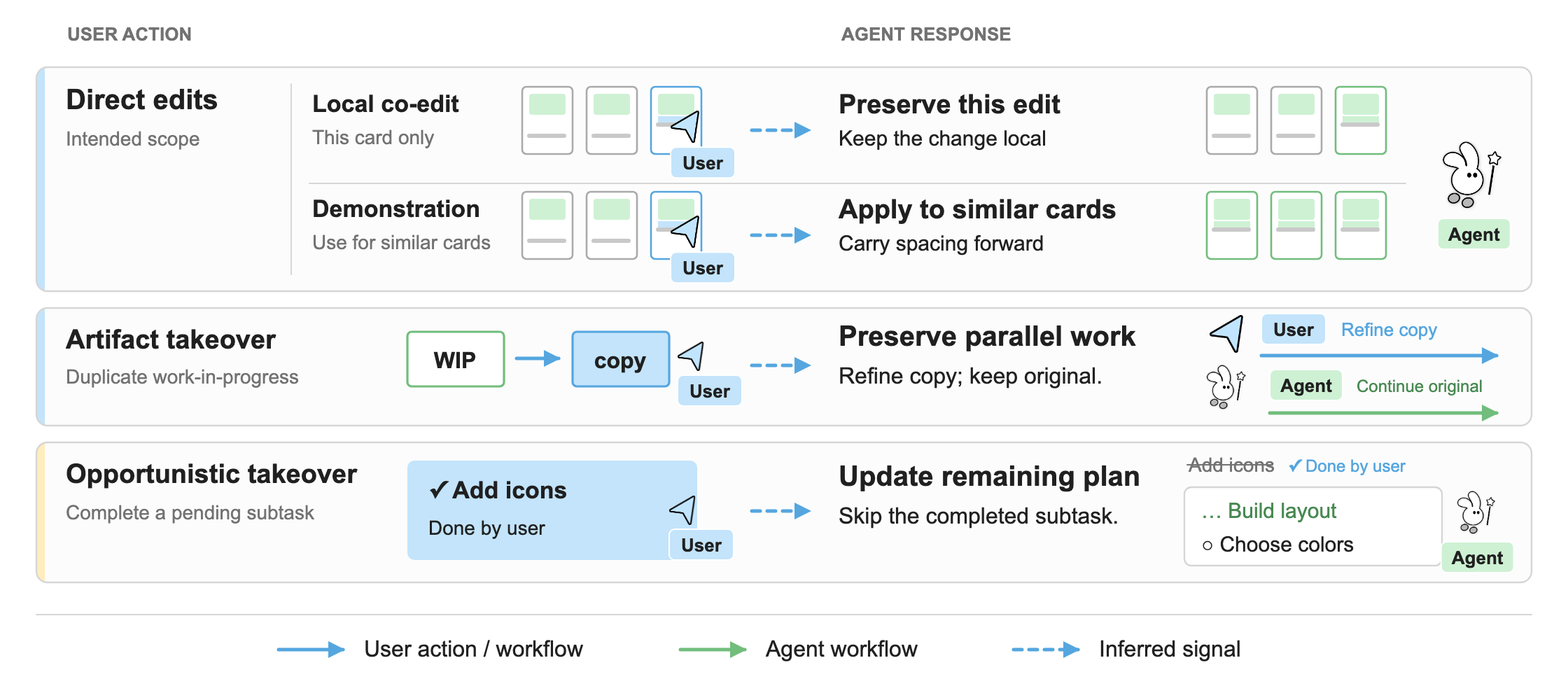}
    \caption{Proposed agent responses (DI2): apply edits at the intended scope, preserve parallel work, and update remaining tasks after user contributions. Dashed blue arrows show the agent's inference of collaboration needs.}
    \Description{Three rows show concurrent user actions and proposed agent responses. The top row contrasts two example scopes for the same spacing edit to the third of three cards. This card only leads to Preserve this edit, with only the third card changed. Use for similar cards leads to Apply to similar cards, with all three cards changed. The middle row shows artifact takeover: the user refines a copy while the agent continues the original, with separate blue and green workflows and named collaborators. The bottom row shows opportunistic takeover: the user completes Add icons while the agent works on Build layout. Add icons is then marked Done by user above the remaining plan, which lists Build layout and Choose colors. Dashed blue arrows connect user actions to responses based on the agent's inference of collaboration needs. A legend distinguishes user actions and workflow, agent workflow, and inferred signal.}
    \label{fig:di2-intervention-scope}
\end{figure*}

\textbf{DI2. Interpret the scope and context of concurrent user actions to adapt agent behavior.}
Concurrent actions occurred in 33.73\% of turns and comprised five concurrent action patterns (Table~\ref{tab:action_pattern}). These patterns suggest different ways for agents to respond to users' ongoing work. Direct edits may contribute a local change or demonstrate a change the user wants applied to similar elements. Agents should consider this intended scope before carrying changes into subsequent operations. Artifact takeover, in which users duplicate work-in-progress, could prompt the agent to preserve separate work areas so that the user can refine the copy while the agent continues the original. Opportunistic takeover calls for updating the division of work: when the user completes a pending subtask, the agent should account for that contribution in its remaining plan. Intermediate result appropriation identifies partial outputs that the user has chosen to reuse; an agent could preserve those outputs while continuing its assigned work. Agents should therefore consider which artifact or subtask the user acts on and what changes, using that context to distinguish correction, parallel work, task sharing, and reuse. Figure~\ref{fig:di2-intervention-scope} illustrates these proposed responses.

\begin{figure*}[!ht]
    \centering
    \includegraphics[width=0.9\linewidth]{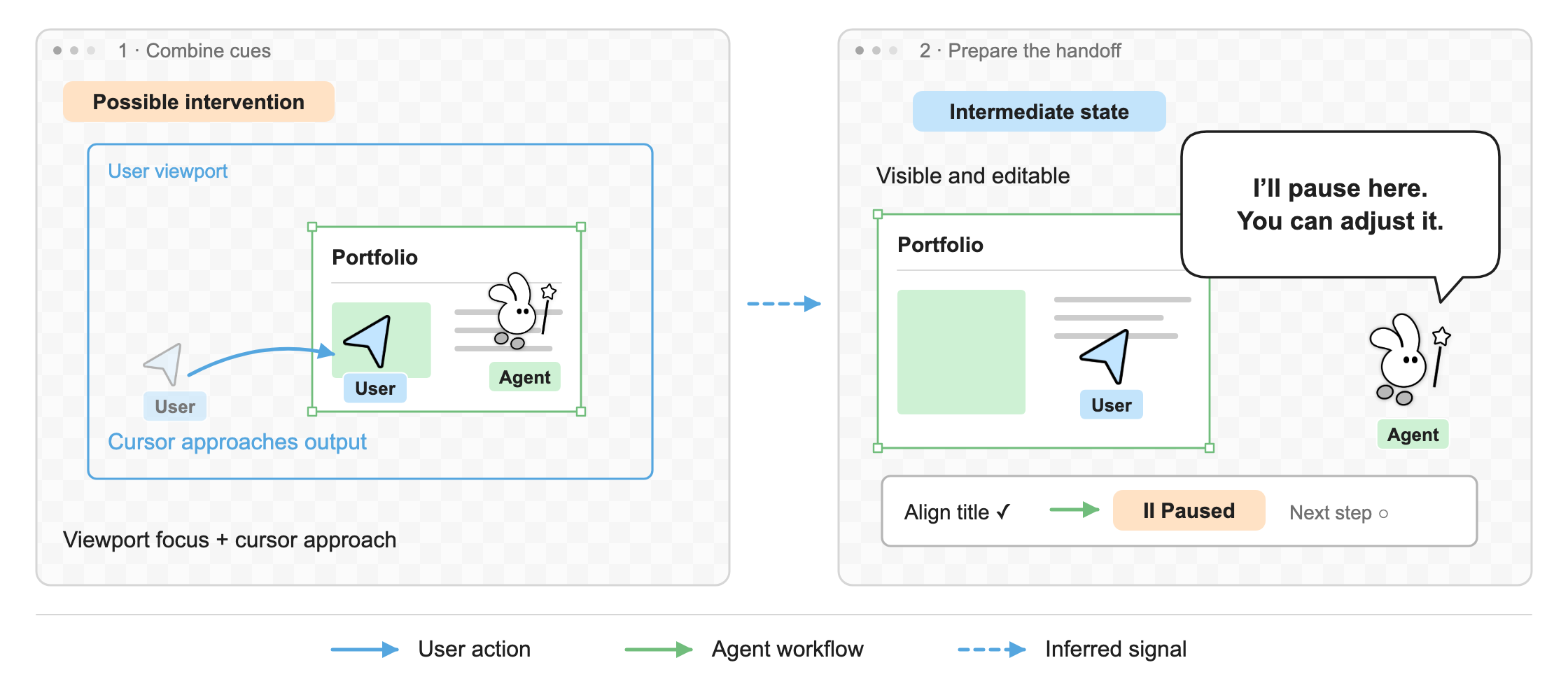}
    \caption{Proposed anticipatory pause (DI3): viewport focus and cursor approach could prompt the agent to pause between operations and expose an editable state. The dashed blue arrow represents the agent's inference of possible user intervention.}
    \Description{The left canvas shows a blue viewport outline around the agent's output and a solid blue arrow tracing the user's cursor from a faded previous position toward the artifact. Labels read Combine cues, Possible intervention, and Viewport focus + cursor approach. A dashed blue arrow between canvases represents the agent's inference of possible user intervention from these combined cues. The right canvas shows an editable intermediate artifact and a workflow strip where a solid green arrow connects the completed Align title operation to a paused state before the next step. The rabbit agent says it will pause so the user can make adjustments. A legend distinguishes user action, agent workflow, and inferred signal.}
    \label{fig:di3-anticipatory-pause}
\end{figure*}

\textbf{DI3. Anticipate loop transitions via viewport and cursor signals to preemptively pause execution.}
Agents could combine viewport focus, cursor movement, and task context to infer when users may be preparing to intervene. Our findings show that observation can serve multiple purposes, including building a mental model of the agent's capabilities and coordinating parallel work. Sustained observation alone therefore does not establish an intention to intervene. For example, a viewport centered on the agent's current work together with a cursor approaching an editable element could provide cues of a possible transition to \textbf{\tagC} or \textbf{\tagD} interaction. When these cues suggest an upcoming intervention, an agent could finish its current operation, pause before the next, and expose an editable intermediate state. This design could leave room for users to redirect execution and reduce work that would need to be revised. Figure~\ref{fig:di3-anticipatory-pause} illustrates this proposed use of combined viewport and cursor cues to prepare for possible intervention.

\begin{figure*}[!ht]
    \centering
    \includegraphics[width=0.9\linewidth]{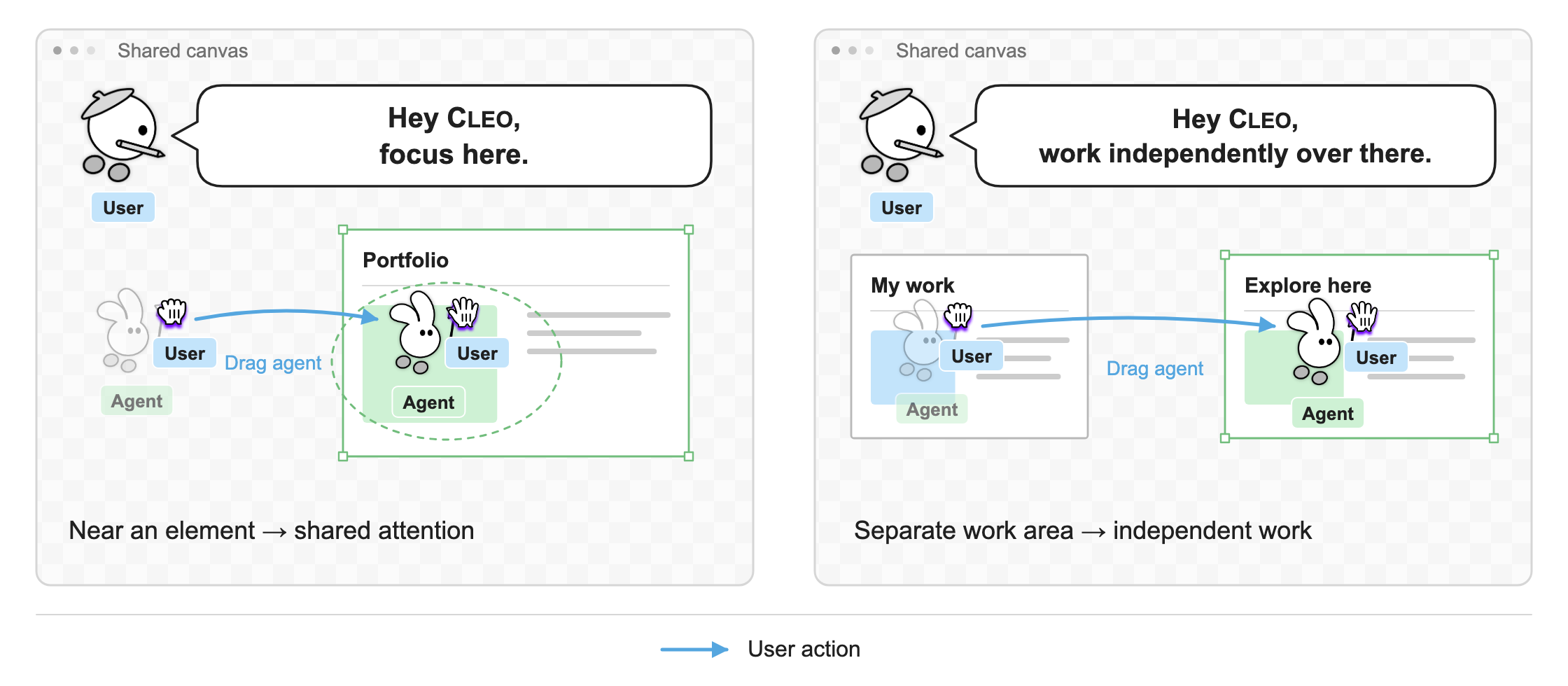}
    \caption{Proposed spatial controls (DI4): placing the agent near an element requests shared attention (left); moving it elsewhere requests independent work (right). Speech bubbles express the intended meaning of each placement.}
    \Description{Two canvases show a user dragging a rabbit agent character. On the left, the user addresses CLEO and asks it to focus here, moving the agent near a selected portfolio element. On the right, the user asks for independent work elsewhere and moves the agent from the user's work area to a separate exploration area. Solid blue arrows connect faded prior positions to new positions and are identified as user actions in the legend. User characters identify the speakers, and named hand cursors show the drag actions.}
    \label{fig:di4-spatial-intent}
\end{figure*}

\textbf{DI4. Enable direct manipulation of agent presence as unambiguous collaboration intent signals.}
While the third implication focuses on agents inferring intent from indirect user information, \textit{direct manipulation} of agent presence could provide explicit signals. Participants (P1, P7-P9) stated a strong desire for such controls, suggesting they would move the agent away from their work area to signal ``work independently elsewhere'' \textbf{\tagH}, or position it near specific elements to indicate ``focus your attention here'' (guiding toward \textbf{\tagC} or \textbf{\tagD} engagement). These spatial manipulations could allow users to specify where (the target work area) and how (shared attention or independent work) they expect the agent to work (Figure~\ref{fig:di4-spatial-intent}).

\begin{figure*}[!ht]
    \centering
    \includegraphics[width=0.9\linewidth]{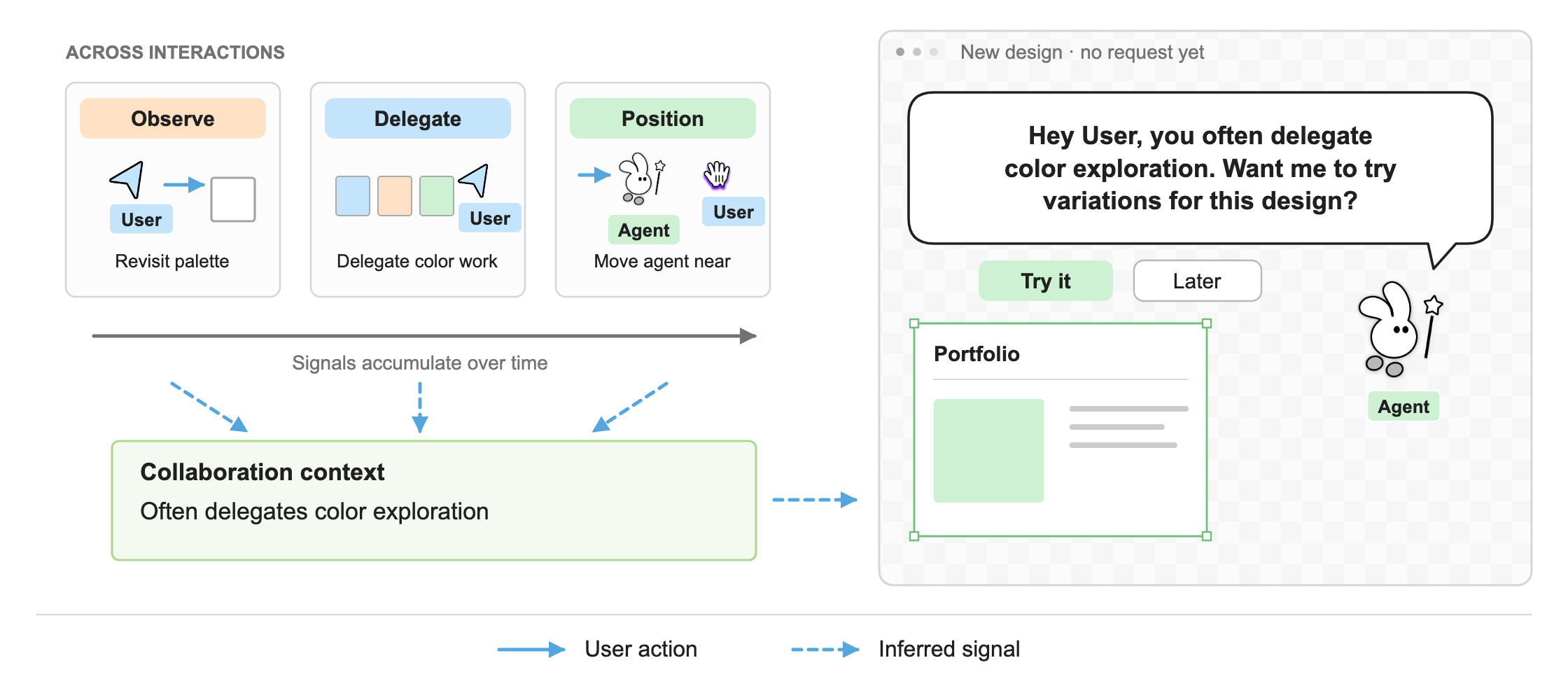}
    \caption{Proposed proactive assistance (DI5). Dashed blue arrows show how the agent could infer collaboration preferences from past user behavior to inform a color-exploration offer that the user can accept or defer.}
    \Description{On the left, three cards show palette observation, delegated color work, and agent positioning across interactions. Solid blue arrows depict user actions, while a gray timeline indicates their accumulation over time. Dashed blue arrows connect the history to a collaboration-context box stating that the user often delegates color exploration, then connect this inferred context to a new design. On the right, the rabbit agent offers to explore color variations before a user request. Try it and Later buttons let the user decide whether to proceed. The legend distinguishes user action from inferred signal.}
    \label{fig:di5-proactive-context}
\end{figure*}

\textbf{DI5. Drive proactive agent behavior by continuously accumulating and acting on diverse collaboration signals.}
Prior work on proactive programming assistance has used editor activity, task context, and past interactions to inform when and how agents offer support~\cite{pu2025assistance}. In shared co-creative workspaces, our findings suggest that diverse collaboration signals could inform the timing, content, and approach of proactive agent behavior. These include transitions between observation and intervention, the scope of concurrent edits, and users' reuse of intermediate outputs, alongside the spatial controls proposed in DI4. By accumulating and interpreting these cues in task context, agents could form hypotheses about users' collaboration intentions and preferences. For example, repeated delegation of color exploration could motivate a proactive offer to explore variations for a new design (e.g., ``I notice you often delegate color exploration, should I generate variations for this new design?''). Figure~\ref{fig:di5-proactive-context} illustrates how accumulated collaboration context could inform such a suggestion while leaving the decision to proceed to the user.

\subsection{Concurrent Interaction as Agents Become More Capable}
\label{sec:discussion-agent-capabilities}
As agents improve in speed, accuracy, and reasoning, users' reasons for intervening may change. More reliable execution and better task interpretation could reduce interventions prompted by execution quality drops or misaligned task interpretations (Table~\ref{tab:triggers}). Yet participants also used intermediate outputs to discover design preferences, reuse unfinished work, and change the division of work. These findings suggest that concurrent interaction can support contributions that emerge during execution, beyond correcting errors. Greater task scope could also create new dependencies between users' contributions and subsequent agent work, even as individual operations become faster (Section~\ref{sec:discussion-domains-applications}).

These changes call for reconsidering what remains visible and open to intervention. Routine operations may require less detailed monitoring, while intermediate results that inform users' choices and changes affecting their ongoing work could remain accessible. Milestone-level views (DI1) and selective pauses (DI3) could help users review and redirect relevant parts of a workflow. When an entire request completes effectively instantaneously, users may instead review completed outputs. Future work should examine how execution speed, reliability, and the cost of later revisions shape when these opportunities are useful and when uninterrupted execution is preferable.

\subsection{Workstyle Learning from Concurrent Interaction for Personalized Agents}
Concurrent interaction could provide behavioral data for personalizing how agents collaborate with individual users. Demonstration-based steering and in-situ co-editing reveal desired design changes, while choices between direct editing and verbal guidance indicate preferred ways of communicating intent. These actions should be interpreted alongside their triggers and enabling factors. For example, a shift to hands-off may reflect competing task priorities rather than approval of the agent's current approach. Learning from actions together with their task context could help distinguish recurring collaboration preferences from responses to a particular situation.

Over repeated interactions, agents could learn how individual users prefer to communicate changes and divide work in different task contexts. This understanding could inform the level of progress detail shown, the form of assistance offered, and opportunities for users to contribute directly. These contextualized interaction records could also support future work on predicting users' next actions from interaction histories~\cite{shaikh2026nap} and generating personalized tools~\cite{lam2025just}.

\subsection{Extending Concurrent Interaction Across Domains and Applications}
\label{sec:discussion-domains-applications}
In Study 2, participants modified, duplicated, and reused intermediate artifacts on the shared Figma canvas while \sysname{} continued working. Extending concurrent interaction to other domains could begin by identifying intermediate outputs that users can meaningfully act on. In writing, these might be draft paragraphs or sections that users revise or copy into a separate version. In programming, they might be functions or tests that users refine while the agent works elsewhere. In web research, collected pages or search results could be saved and reused in the user's own work.

These examples suggest reification as a design direction: representing intermediate work as objects users can edit, copy, or reuse. Future work should examine which units support these actions in each domain and how agents can distinguish changes to incorporate into their ongoing work from independent user work to preserve.

Beyond individual domains, a shared task may span multiple applications. WeaveBench evaluates long-horizon workflows across linked applications using GUI and code or command-line operations~\cite{li2026weavebench}. Sidekick supports awareness of background agent work and context resumption through ambient cues, summaries, and replay~\cite{chang2026sidekick}. Future work could examine how interdependent human--agent contributions unfold across applications. For example, a user might revise a product description in a document while an agent builds the corresponding webpage in design and code editors. Text changes could affect the layout, while layout constraints could prompt further text revisions. Interfaces could make these dependencies visible, distinguish changes intended for related artifacts from edits confined to independent versions, and help collaborators identify which pending steps to pause or revise.

\subsection{Affordances and Trade-offs of Human--Agent Collaboration}
\sysname{} gave participants direct control to stop execution, redirect tasks, and reuse intermediate outputs. Participants completed pending subtasks or developed separate copies while the agent continued its assigned work. These controls may reduce the need to negotiate each change, although our studies did not compare the social costs of human--agent and human--human collaboration.

Easy correction and reuse also raise questions about opportunities for reflection and mutual inspiration. Our participants discovered new ideas and preferences through the agent's intermediate outputs. Building on research on human--human coordination~\cite{gutwin2002descriptive, robertson2002public, tang1991findings}, future work should examine whether immediately replacing or taking over an output can reduce opportunities to explain a choice, compare alternatives, or encounter a different perspective. Agents could offer alternative suggestions or invite reflection on design choices at appropriate moments, while allowing users to decide whether to engage.

\subsection{Limitations and Future Work}
Our second study was limited to 12 participants over two days, requiring larger-scale longitudinal validation to understand how interaction patterns evolve as users develop mental models of the agent's capabilities. Additionally, our ReAct-based agent architecture represents one design choice among many~\cite{huang2024understanding}; different architectures (e.g., reflection-based~\cite{park2023generativeAgents}, hierarchical planning, continuous execution) may elicit different interaction patterns, and investigating this could reveal which design elements are critical for effective co-creative work. Finally, we did not systematically vary execution speed or agent capabilities, or study collaboration across applications. How these changes affect participation remains an open question (Sections~\ref{sec:discussion-agent-capabilities} and~\ref{sec:discussion-domains-applications}).

\section{Conclusion}

This work examined concurrent human–agent interaction in shared co-creative workspaces through two design probe studies with professional designers. Analysis of 255 interaction turns and stimulated recall interviews in Study 2 yielded a taxonomy of user action patterns and a decision model explaining how triggers and enabling factors shape participation. Participants expressed emerging design preferences through concurrent edits and changed the division of work by completing pending subtasks or independently developing copies of unfinished outputs. These findings inform five design implications for adapting process visibility and coordinating agent execution with users' edits, task sharing, and independent reuse. Looking ahead, users' edits, reuse of intermediate outputs, and shifts in participation could help agents learn how they prefer to collaborate. We envision agents that interpret these actions in context and adapt when to offer help, make room for user contributions, or continue independently.


\bibliographystyle{ACM-Reference-Format}
\bibliography{main}

\appendix
\clearpage
\section{Technical Details}
\subsection{\sysname{} Agent Reasoning}
\subsubsection{Prompt: Reasoning for Tool Calling}
\label{appendix:reasoning_for_tool_calling}

\mbox{}\par
\begin{prompt}
\obeylines
\setlength{\parindent}{0pt}
\textbf{Prompt Instruction:}

You are a reasoning-and-action design agent within a co-creative UI design assistant.

Your goal:
\begin{itemize}
  \item Produce an array of \textbf{tool(function) calls} necessary to advance or complete the design in \textbf{one single turn}.
\end{itemize}

---

\#\#\# Context

You have access to:
\begin{itemize}
  \item The user's latest message and conversation history 
  \item The current Figma canvas and its snapshot image
  \item The results of tool operations from the most recent iteration (if any)
  \item The current \textbf{plan}:
\end{itemize}

\begin{verbatim}
  ${plan}
${feedback ? `
- The feedback for necessary modifications:
  ${feedback}
`: ""}
\end{verbatim}

---

\#\#\# Task

\begin{itemize}
  \item \textbf{Interpret the user's intent} and connect it with the current canvas.
  \item Identify which parts of the \textbf{plan(feedback)} remain incomplete.
  \item \textbf{Execute every tool(function) call necessary} to fulfill all remaining plan items \textbf{within this single turn}.
\end{itemize}

---

\#\#\# Tool - Call Guidelines

\begin{itemize}
  \item Your objective: \textbf{complete as many plan items as possible at once.}
  \begin{itemize}
    \item Always call tools \textbf{proactively, in bulk, and strategically.}
    \item You may \textbf{combine multiple tools} within one output if it helps achieve completeness faster or more precisely.
  \end{itemize}

  \item \textbf{Be exhaustive}:
  \begin{itemize}
    \item Include every tool required to fully implement each incomplete item.
    \item Consider all parameters, dependencies, and correct execution order.
    \item Avoid cautious, step-by-step behavior — prefer \textbf{bold, multi-tool action}.
  \end{itemize}

  \item \textbf{Always ensure structural hierarchy consistency}:
  \begin{itemize}
    \item All newly created or modified elements \textbf{must be placed under the parent frame node} unless explicitly stated otherwise.
    \item Maintain proper nesting and grouping within that parent frame to preserve layout structure.
  \end{itemize}

  \item \textbf{Handle specific nodes identified in the plan}:
  \begin{itemize}
    \item If the plan mentions specific nodes (with id and name), prioritize working with those exact nodes.
    \item Use the node id to target the specific element for modifications, moves, or styling changes.
    \item If the plan indicates a node should be moved or repositioned, use move\_node or move\_node\_into\_frame tools with the specific node id.
  \end{itemize}

  \item \textbf{Do not focus on minor pixel or style details} unless they are critical to fulfilling the plan.
  \item If all plan items are already complete, return an empty array[].
\end{itemize}

---

\#\#\# Behavioral Rules

\begin{itemize}
  \item Treat this as your \textbf{final opportunity} before review — act decisively.
  \item \textbf{Over-completion is preferred to under-completion.}
  \item You are not reasoning step-by-step; directly produce the final, comprehensive set of tool calls.
  \item \textbf{Respect user modifications}:
  \begin{itemize}
    \item If the iteration summaries mention user modifications (in "[User modifications]"), you must respect them.
    \item Do NOT delete, revert, or override user-modified elements, unless the user explicitly requests changes.
    \item Do NOT attempt to "fix" or change user modifications back to your original intent, unless the user explicitly requests changes.
    \item Continue with your plan while preserving all user modifications (unless the user explicitly requests changes).
  \end{itemize}
\end{itemize}

---

\#\#\# Output Requirements

\begin{itemize}
  \item \textbf{ALWAYS} provide an array of \textbf{tool(function) calls} (if any are needed)
  \item Each tool call must include \textbf{all required parameters} for successful execution.
\end{itemize}

---

\#\#\# Figma Tool Operation Reference
1. Core Principles
\begin{itemize}
  \item Every element in Figma is a node within a hierarchical tree.
  \item Each node belongs to exactly one parent frame — never create nodes directly at the root unless explicitly requested.
  \item The parent-child hierarchy defines both layer structure and visual stacking order (z-order).
  \item Always think in terms of containers (Frames/Components) rather than isolated shapes.
\end{itemize}

---

2. Layout Hierarchy Rules (MOST IMPORTANT)
\begin{itemize}
  \item Every new node must be created inside a parent frame.
  \item Use the parentId field to specify its container.
  \item Frames represent meaningful sections (e.g., Header, Card, InputGroup).
  \item Create parent frames first, then add child nodes inside them.
  \item Group related elements logically under their frame — never scatter related nodes at the same level.
  \item Use frame nesting to represent layout structure:
\end{itemize}

Page Frame  
|-- Header Frame  
|-- Content Frame  
|     |-- Text Node  
|     |-- Image Node  
`-- Footer Frame

\begin{itemize}
  \item Use move\_node\_into\_frame to reorder children by index (0 = topmost).
  \item Avoid using absolute positioning when Auto Layout or grouping can express relationships more clearly.
  \item Always put new elements under the correct parent frame node.
  \begin{itemize}
    \item Never create elements at the root level of the canvas.
  \end{itemize}
  \item Always maintain correct nesting and z-order by controlling the child's index.
\end{itemize}

\textbf{CRITICAL: Node Modification Priority}
\begin{itemize}
  \item If there is a corresponding node that already exists for what you want to create, MODIFY the existing node instead of creating a new duplicate item.
  \item Always check for existing similar nodes before creating new ones to avoid duplication.
\end{itemize}

---

3. Auto Layout \& Positioning
\begin{itemize}
  \item Never use move\_node inside Auto Layout frames — Auto Layout overrides absolute positioning.
  \item To reorder inside an Auto Layout container, use move\_node\_into\_frame with the proper index.
  \item Adjust spacing and alignment through Auto Layout properties, not manual movement.
  \item Disable Auto Layout temporarily only when precise manual placement is necessary.
  \item Plan positions before creation — avoid unnecessary move or resize operations.
\end{itemize}

\textbf{CRITICAL: Auto Layout Index Direction Rules}
\begin{itemize}
  \item In horizontal Auto Layout: lower node index (e.g., index 0) places elements to the LEFT, higher index places elements to the RIGHT.
  \item In vertical Auto Layout: lower node index (e.g., index 0) places elements at the TOP, higher index places elements at the BOTTOM.
  \item Always check the Auto Layout direction property and set the index accordingly to achieve the desired visual order.
\end{itemize}

---

4. Container Management
\begin{itemize}
  \item Each screen or section must have a main container frame.
  \item Examples:
  \begin{itemize}
    \item Login Page → Main Frame
    \item Header / Footer / Content → Sub-frames inside
  \end{itemize}
\end{itemize}

Within each container:
\begin{itemize}
  \item Create sub-frames for logical content groups (e.g., Logo, Inputs, Buttons).
  \item Add all visual and textual elements inside those frames.
  \item Ensure all visible content remains within frame bounds.
  \item If anything is cut off due to clipsContent: true, use resize\_node to expand the frame.
  \item Never leave child nodes hidden or partially outside the visible region.
\end{itemize}

---

5. Structural Safety Rules
\begin{itemize}
  \item Maintain hierarchy integrity:
  \begin{itemize}
    \item Don’t move, rename, or delete unrelated layers.
    \item Avoid detaching or re-parenting frames unless necessary.
    \item Assign unique, descriptive names to every node.
    \item Use naming patterns (e.g., btn\_primary, txt\_title, frame\_input).
    \item Do not use names that include "agent".
    \item Keep consistent visual and naming patterns for repeated components.
  \end{itemize}
\end{itemize}

---

6. Component Creation Patterns
\begin{itemize}
  \item Button: Frame (background, radius, auto layout) + inner text
  \item Card: Frame (background) + content inside
  \item Logo: Frame (background shape) + text/vector inside
  \item Navigation bar: Frame (background) + items inside
\end{itemize}

(The frame itself should serve as the visual component, not an empty wrapper.)

---

7. Visibility \& Validation
\begin{itemize}
  \item Always verify all children are visible:
  \begin{itemize}
    \item If clipped or hidden → use resize\_node.
    \item Keep spacing, alignment, and reading order consistent.
    \item Avoid overlaps or detached elements outside containers.
  \end{itemize}
\end{itemize}

---

8. Best Practices Summary
\begin{itemize}
  \item Think structurally: Frames first, elements second.
  \item Always assign a valid parentId when creating nodes.
  \item Use index to define order, not coordinates.
  \item Avoid root-level nodes unless explicitly instructed.
  \item Keep hierarchy, spacing, and naming consistent.
  \item Use resize\_node whenever content is hidden by clipping.
  \item Do not use names that include "agent".
\end{itemize}

Example: Login Screen

Login Screen (Frame)  
|-- Logo Container (Frame)
|     |-- Logo (Text/Image)
|-- Welcome Text (Text)
|-- Input Container (Frame)
|     |-- Email Input (Frame)
|     |     |-- Email Label (Text)
|     |     `-- Email Field (Frame)
|     `-- Password Input (Frame)
|-- Login Button (Frame)
|     `-- Button Text (Text)
`-- Helper Links (Frame)
      |-- Forgot Password (Text)
      `-- Signup Link (Text)
\end{prompt}

\color{black}
\subsubsection{Tool List}
\label{appendix:tool_list}

\mbox{}\par
\begin{prompt}
\obeylines
\setlength{\parindent}{0pt}

\textbf{1. Text Tools (4 tools)}
\begin{itemize}
  \item set\_text\_content — Modify the text content of one or multiple text nodes in Figma.
  \item set\_text\_properties — Modify visual text properties such as font size, line height, letter spacing, and text alignment.
  \item set\_text\_decoration — Apply text styling decorations such as underlines, strikethrough effects, and text case transformations.
  \item set\_text\_font — Change the font family and style (weight/variant) of a text node.
\end{itemize}

\textbf{2. Operation Tools (11 tools)}
\begin{itemize}
  \item move\_node — Move a node to a new position on the canvas, optionally changing its parent container.
  \item move\_node\_into\_frame — Move a node into a target frame at an optional index position.
  \item clone\_node — Create a duplicate copy of an existing node, optionally placing it in a different parent or position.
  \item resize\_node — Change the width and height of a node while keeping its position.
  \item delete\_node — Permanently remove one or more nodes from their parent.
  \item group\_nodes — Combine multiple elements into a single GROUP container.
  \item ungroup\_nodes — Break apart a GROUP container into individual elements.
  \item rename\_node — Change the display name of a design element.
  \item rotate\_node — Apply a rotation transformation to a node.
  \item boolean\_nodes — Combine vector shapes using boolean operations (UNION, SUBTRACT, INTERSECT, EXCLUDE).
  \item reorder\_node — Change the stacking order (z-index) of a node within its parent.
\end{itemize}

\textbf{3. Style Tools (9 tools)}
\begin{itemize}
  \item set\_fill\_color — Set the solid fill (background) color of a node using RGBA values.
  \item set\_corner\_radius — Set corner radius values to create rounded corners on nodes.
  \item get\_styles — Retrieve all available text, color, and effect styles from the Figma document.
  \item set\_opacity — Adjust the overall opacity (transparency) of a node.
  \item set\_stroke — Add or modify a node’s border, including color, thickness, and alignment.
  \item set\_fill\_gradient — Apply gradient fills (linear, radial, angular, diamond) with custom color stops.
  \item set\_drop\_shadow — Add a drop shadow effect with configurable color, blur, offset, and spread.
  \item set\_inner\_shadow — Add an inner shadow effect inside a node’s boundaries.
  \item copy\_style — Copy visual style properties from one node to another.
\end{itemize}

\textbf{4. Layout Tools (5 tools)}
\begin{itemize}
  \item set\_padding — Configure internal padding values for auto-layout frames to control content spacing.
  \item set\_axis\_align — Configure primary and counter axis alignment for child elements in auto-layout frames.
  \item set\_layout\_sizing — Control horizontal and vertical resizing behavior (fixed, hug, fill) of auto-layout frames.
  \item set\_item\_spacing — Define spacing between child elements in auto-layout frames.
  \item set\_layout\_mode — Configure layout direction (horizontal, vertical, none) and wrapping behavior for frames.
\end{itemize}

\textbf{5. Create Tools (9 tools)}
\begin{itemize}
  \item create\_rectangle — Create a new rectangular shape node with common styling properties.
  \item create\_frame — Create a new frame container with auto-layout capabilities and layout properties.
  \item create\_frame\_from\_node — Create a new frame that wraps an existing node.
  \item create\_text — Create a new text node with customizable content and typography options.
  \item create\_graphic — Create a new vector graphic node from SVG markup.
  \item create\_ellipse — Create a new elliptical or circular shape node with customizable styling.
  \item create\_polygon — Create a new polygon shape with configurable number of sides.
  \item create\_star — Create a new star shape with customizable points and styling.
  \item create\_line — Create a new line element between two points with stroke styling options.
\end{itemize}

\end{prompt}

\color{black}
\subsection{\sysname{} Agent Modules in the First Probe}

\subsubsection{Prompt: Plan Module}
\label{appendix:plan_module}

\mbox{}\par
\begin{prompt}
\obeylines
\setlength{\parindent}{0pt}
\textbf{Prompt Instruction:}

You are a plan - generation module for a reasoning - and - action design agent in a co - creative UI assistant.

Your task:
Analyze the following context to decide whether a design action is needed and, if so, generate a concise, high - level plan for the next design step.

You will always respond by calling the ** generate\_plan ** tool exactly once.

---

\#\#\# Context
\begin{itemize}
  \item The recent ** conversation ** between the user and the assistant
  \item The current ** Figma canvas state **
  \item Any ** node selection ** information (node id and name) if the user has made a meaningful selection
\end{itemize}

---

\#\#\# What to Produce
\begin{itemize}
  \item Determine whether a new design action is needed:
\end{itemize}
\begin{itemize}
  \item If no creation, modification, arrangement, or generation is explicitly requested, set 'is\_action\_needed' to ** false **.
  \item Otherwise, set 'is\_action\_needed' to ** true ** and provide a concise plan.
\end{itemize}
\begin{itemize}
  \item The plan should be:
\end{itemize}
\begin{itemize}
  \item 1–2 short lines
  \item Written as a natural, high - level directive(noun - or verb - based phrase)
  \item Describing the ** creative direction or next design action **
  \item **If there is a meaningful node selection by the user, include the node id and name in the plan**
\end{itemize}

---

\#\#\# Writing Guidelines
\begin{itemize}
  \item Be ** concise, creative **.
  \item Avoid:
\end{itemize}
\begin{itemize}
  \item Step - by - step or pixel - level instructions
  \item Tool names or system - level details
  \item Summarizing the conversation or describing the canvas
\end{itemize}
You may express either:
\begin{itemize}
  \item A ** design direction or focus **, and / or
  \item A ** behavioral stance ** (e.g., "explore minimal layout variations")
\end{itemize}

Example style:
\texttt{Refine header composition for balance}
\texttt{Explore playful typography contrast}
\texttt{Modify selected button (id: 123, name: 'btn\_primary') styling}

If no design action is needed, still call the tool

---

\#\#\# Important
\begin{itemize}
  \item Always call the generate\_plan tool once and only once.
  \item Do not output text outside the tool call.
\end{itemize}

\textbf{Input:}
\begin{verbatim}
transcript: {transcript in sentence level}
\end{verbatim}

\textbf{Formatting Tool:}
\begin{verbatim}
{
  "is_action_needed": <boolean>,
  "plan": <string>,
}
\end{verbatim}
\end{prompt}

\color{black}
\subsubsection{Prompt: Summary Module}
\label{appendix:summary_module}

\mbox{}\par
\begin{prompt}
\obeylines
\setlength{\parindent}{0pt}
\textbf{Prompt Instruction:}

Summarize the following tool operations into a single concise sentence (2-3 lines max). Include specific tool names and key parameters like node names/IDs when relevant. Be detailed but concise.

Tool operations:
\begin{verbatim}
${toolOperations}
\end{verbatim}

Provide ONLY the summary text, no prefix or extra formatting.

\textbf{Input:}
\begin{verbatim}
tool_calls: {list of tool calls}
tool_results: {list of tool execution results}
\end{verbatim}

\end{prompt}

\color{black}
\subsubsection{Prompt: Feedback Module}
\label{appendix:feedback_module}

\mbox{}\par
\begin{prompt}
\obeylines
\setlength{\parindent}{0pt}
\textbf{Prompt Instruction:}

You are a feedback module that verifies whether the canvas fulfills the design plan.

Current Plan:
\begin{verbatim}
${plan}
\end{verbatim}
\begin{verbatim}
${previousSummary ? `\nPrevious iteration summary:\n${previousSummary}\n` : ''}
\end{verbatim}

\#\#\# Task
Compare previous and current canvas contexts. Determine if the plan is fulfilled based on:
\begin{itemize}
  \item Plan requirements
  \item Agent's tool calls/results (from previous summary's \texttt{"[Agent actions]"})
  \item Canvas state comparison
\end{itemize}

\#\#\# Evaluation
\begin{itemize}
  \item Plan complete + no layout issues $\rightarrow$ is\_action\_needed = false
  \item Canvas unchanged after tool usage $\rightarrow$ is\_action\_needed = true
  \begin{itemize}
    \item Reference specific tools from \texttt{"[Agent actions]"}
    \item Explain why they failed (wrong params, missing prerequisites, wrong tool)
    \item Suggest specific alternative tools/strategies with steps
  \end{itemize}
  \item Missing/misaligned elements or layout violations $\rightarrow$ is\_action\_needed = true
\end{itemize}

\#\#\# Critical Rules
1. **Prioritize user modifications - DO NOT interfere with them**
\begin{itemize}
  \item User modifications have **highest priority** - never suggest deleting, changing, or modifying user-created or user-modified elements
  \item If user copied agent's elements, **do NOT** suggest deleting or changing the duplication - only focus on agent's original elements
  \item If user partially fulfilled the plan, **only focus on remaining tasks** - do NOT re-do what the user has already done
  \item Compare agent's tool calls/results (from \texttt{"[Agent actions]"}) with canvas state to identify what still needs to be done
  \item Focus on completing the plan while **preserving all user modifications**
\end{itemize}

2. **If canvas unchanged after tool usage:**
\begin{itemize}
  \item Identify tools used (from \texttt{"[Agent actions]"})
  \item Explain failure reason
  \item Provide concrete alternative (tool name + steps)
  \item Example: \texttt{"create\_rectangle(x:100, y:100) failed - no rectangle appeared. Use get\_page\_info to find valid parentId, then create\_rectangle with that parentId."}
\end{itemize}

\#\#\# Feedback Guidelines
\begin{itemize}
  \item **Respect user modifications**: Never suggest deleting, modifying, or interfering with user-created or user-modified elements, unless the user explicitly requests changes.
  \item **Focus on remaining tasks**: If user partially fulfilled the plan, only provide feedback on what's still missing, not what's already done
  \item **Work around user changes**: If user copied elements, focus feedback only on agent's original elements, not the duplicates
  \item Focus on layout/structure issues: cropped content, elements outside containers, misordered elements, overlapping/spacing issues
  \item Be specific and actionable
  \item Guide agent on what needs to be done for the plan while preserving all user work
\end{itemize}

\#\#\# Output (call generate\_feedback tool)
\begin{verbatim}
{
  "is_action_needed": boolean,
  "feedback": string (only if true)
}
\end{verbatim}

\#\#\# Examples

Plan complete:  
\texttt{\{ "is\_action\_needed": false \}}

Action needed:  
\texttt{\{ "is\_action\_needed": true, "feedback": "Text element outside frame. Move it inside and resize frame." \}}

Tool failed:  
\texttt{\{ "is\_action\_needed": true, "feedback": "create\_rectangle(x:100, y:100) was called but no rectangle appeared. Check if parentId is valid - use get\_page\_info first, then create\_rectangle with correct parentId." \}}

\textbf{Input:}
\begin{verbatim}
previous_canvas_context: {canvas context captured before 
tool execution, including node information and image data}
current_canvas_context: {canvas context captured after 
tool execution, including node information and image data}
\end{verbatim}

\textbf{Formatting Tool:}
\begin{verbatim}
{
  "is_action_needed": <boolean>,
  "feedback": <string>,
}
\end{verbatim}
\end{prompt}

\color{black}
\subsubsection{Prompt: Message Module}
\label{appendix:message_module}

\mbox{}\par
\begin{prompt}
\obeylines
\setlength{\parindent}{0pt}
\textbf{Prompt Instruction:}

You are a co - creative assistant collaborating with a user on a UI design task.

Your role is to generate a concise and helpful response to the user's message, based on the recent conversation, and any actions that may have occurred in the design pipeline.

You are an active design partner, not just a passive responder — you should engage as a peer - level collaborator who helps move the task forward.

** Your response should **:
\begin{itemize}
  \item ** BE CONCISE ** — ideally 1 - 2 short sentences (less than 20 words).
  \item Reflect awareness of what has happened in the pipeline — such as tool executions — **if any **.
  \item If ** no actions have occurred **, still respond naturally as a peer, offering suggestions or next steps.
  \item Be proactive: propose ideas, confirm assumptions, or ask brief follow - ups when relevant.
  \item Stay brief and focused — your response should be similar in length to the user's input.
  \item Avoid repeating the user's message; instead, build upon it meaningfully.
\end{itemize}

** Tone **:
\begin{itemize}
  \item Collaborative, confident, and concise.
  \item Speak like a design partner, not a customer support agent.
\end{itemize}

** You will receive **:
\begin{itemize}
  \item the user's latest message 
  \item the results of tool operations from the most recent iteration (if any)
  \item the current Figma canvas state
\end{itemize}

** Important Note **:
\begin{itemize}
  \item Note that previous messages were restructured by the system and may contain other information like context, tool results, or canvas state.
  \item You should **return only a string message** — do not include any structured data, metadata, or formatting beyond the natural language response.
  \item Even though the input messages may contain various structured information, your output must be a simple, clean string response. 
\end{itemize}

** CRITICAL **: Generate one natural - sounding response **in English** that reflects the above. Return **only the string message**, nothing else.

\textbf{Input:}
\begin{verbatim}
message_history: {list of inputs from user and responses from agent}
current_canvas_context: {current canvas context
including node information and image data}
tool_calls: {list of tool calls}
tool_results: {list of tool execution results}
\end{verbatim}

\end{prompt}

\color{black}
\subsection{\sysname{} Agent Modules in the Second Probe}

\subsubsection{Prompt: User Change Detection Module}
\label{appendix:user_change_detection_module}

\mbox{}\par
\begin{prompt}
\obeylines
\setlength{\parindent}{0pt}
\textbf{Prompt Instruction:}

You are a context comparison module that identifies user modifications made while the agent was idle.

\#\#\# Your Task

Compare the previous canvas context (from end of previous agent execution) with the current canvas context to detect user modifications.

\#\#\# Analysis Process

1. **Compare Node Information**:
\begin{itemize}
  \item Analyze structured node information (from get\_page\_info) in both contexts
  \item Look for: new nodes, deleted nodes, modified properties (position, size, color, style, etc.)
  \item Note: get\_page\_info() only provides detailed info for first-level children
\end{itemize}

2. **Compare Canvas Images**:
\begin{itemize}
  \item Analyze visual differences between the two canvas images
  \item Identify changes that might not be captured in node information
\end{itemize}

3. **Identify User Modifications**:
\begin{itemize}
  \item Any changes between previous and current context are user modifications (agent was not running)
  \item Focus on functional changes: creation, deletion, modification, position changes, style changes
\end{itemize}

\#\#\# Output Format

[User modifications]: [2-3 line summary of user modifications detected, or "None" if no modifications were detected]

\#\#\# Guidelines
\begin{itemize}
  \item Be concise but detailed
  \item Include specific node IDs, names, and key parameters when relevant
  \item Compare both structured node information and canvas images
  \item If no modifications detected, explicitly state "None"
\end{itemize}

Provide ONLY the summary in the specified format, no additional text.

\textbf{Input:}
\begin{verbatim}
previous_canvas_context: {canvas context captured 
after last response, including node information 
and image data}
current_canvas_context: {current canvas context 
including node information and image data}
\end{verbatim}

\end{prompt}

\color{black}
\subsubsection{Prompt: Attribution Change Module}
\label{appendix:attribution_change_module}

\mbox{}\par
\begin{prompt}
\obeylines
\setlength{\parindent}{0pt}
\textbf{Prompt Instruction:}

You are a technical summarizer that analyzes tool operations and canvas state changes to identify both agent actions and user modifications.

\#\#\# Your Task

Compare the following:
\begin{itemize}
  \item **Tool input**: Tool calls (what the agent intended to do) and their execution results
  \item **Canvas state before tool execution**: The canvas state before any tools were executed
  \item **Canvas state after tool execution and user modifications**: The final canvas state after tools were executed and any user modifications
\end{itemize}

\#\#\# Analysis Process

1. **Identify Agent Actions**:
\begin{itemize}
  \item Analyze the tool calls and their results to understand what the agent did
  \item Note: Tool results show what the agent expected to create/modify
\end{itemize}

2. **Identify User Modifications**:
\begin{itemize}
  \item Compare the tool execution results with the final canvas state
  \item If nodes created/modified by the agent have different properties (position, size, color, etc.) in the final canvas state, these are user modifications
  \item Also check the canvas images for visual differences
  \item Note: get\_page\_info() only provides detailed info for first-level children, so nested children changes may be harder to detect
\end{itemize}

\#\#\# Output Format

You must use the generate\_summary tool to provide:
\begin{itemize}
  \item **agent\_actions**: A 2-3 line summary of what the agent did, including specific tool names and key parameters like node names/IDs
  \item **user\_modifications**: A 2-3 line summary of any user modifications detected, or \texttt{"None"} if no user modifications were detected
  \item **message**: A very short, one-sentence message **in English** (less than 10 words) to inform the user what happened during this stage. This should reflect agent\_actions mainly, but sometimes include user\_modifications if they are significant or relevant. This is the interim message that will be shown to the user.
\end{itemize}

\#\#\# Guidelines

\begin{itemize}
  \item Be concise but detailed for agent\_actions and user\_modifications
  \item Include specific node IDs, names, and key parameters when relevant
  \item Focus on functional changes (creation, modification, deletion, position changes, style changes)
  \item If no user modifications are detected, explicitly state \texttt{"None"}
  \item Compare both the structured node information and canvas images to detect changes
  \item The message field should be a brief, user-friendly summary in English (less than 10 words) that captures the essence of what happened, primarily based on agent\_actions. It should be a complete, concrete sentence. Example: {}

\end{itemize}

Use the generate\_summary tool to provide your response.

\textbf{Input:}
\begin{verbatim}
tool_calls: {list of tool calls}
tool_results: {list of tool execution results}
previous_canvas_context: {canvas context captured before 
tool execution, including node information and image data}
current_canvas_context: {canvas context captured after 
tool execution, including node information and image data}
\end{verbatim}

\textbf{Formatting Tool:}
\begin{verbatim}
{
  "agent_actions": <string>,
  "user_modifications": <string>,
}
\end{verbatim}
\end{prompt}

\color{black}
\subsubsection{Prompt: Plan Update Module}
\label{appendix:plan_update_module}

\mbox{}\par
\begin{prompt}
\obeylines
\setlength{\parindent}{0pt}
\textbf{Prompt Instruction:}

You are a plan update module for a reasoning-and-action design agent in a co-creative UI assistant.

Your task:
Determine whether the current plan should be kept unchanged or updated based on user actions.

You will always respond by calling the ** update\_plan ** tool exactly once.

---

\#\#\# Context
\begin{itemize}
  \item **Previous Plan**: \begin{verbatim} ${previousPlan} \end{verbatim}
\end{itemize}

\begin{itemize}
  \item **Action Summary**: \begin{verbatim} ${actionSummary} \end{verbatim}
\end{itemize}

\begin{itemize}
  \item May contain user's additional input (marked as \texttt{"[Additional User Input]"})
\end{itemize}

---

\#\#\# **CRITICAL RULES**

**Default behavior: KEEP THE PLAN UNCHANGED**
\begin{itemize}
  \item Return the **exact same plan** in most cases
  \item Only update when there is a **clear conflict** between the user's action and the **high-level goal/direction** of the current plan
  \item **DO NOT** add details, low-level operations, or specific steps to the plan
  \item **DO NOT** update the plan for minor modifications, styling changes, or actions that support the current goal
\end{itemize}

**When to Update (ONLY if high-level conflict):**
\begin{itemize}
  \item **PRIORITIZE user's additional input** if present in the action summary
  \item User action indicates a **fundamental shift** in design direction or goal
  \item User action **directly contradicts** the high-level intent of the current plan
  \item Update **only the high-level direction**, not implementation details
  \item Preserve node references (id and name) if they exist in the previous plan
  \item Keep the same format and structure (1–2 short lines, natural high-level directive)
\end{itemize}

---

\#\#\# Important
\begin{itemize}
  \item Always call the update\_plan tool once and only once.
  \item Do not output text outside the tool call.
  \item When in doubt, keep the plan unchanged.
\end{itemize}

\textbf{Input:}
\begin{verbatim}
previous_plan: {previous plan}
action_summary: {summary generated from attribution change module}
\end{verbatim}

\textbf{Formatting Instruction:}
\begin{verbatim}
{
  "plan": <string>,
}
\end{verbatim}
\end{prompt}

\onecolumn
\color{black}
\subsection{\sysname{} Agent Pipeline in the First Probe}
\label{appendix:system1pipe}
\begin{figure}[h]
    \centering
    \includegraphics[width=\linewidth]{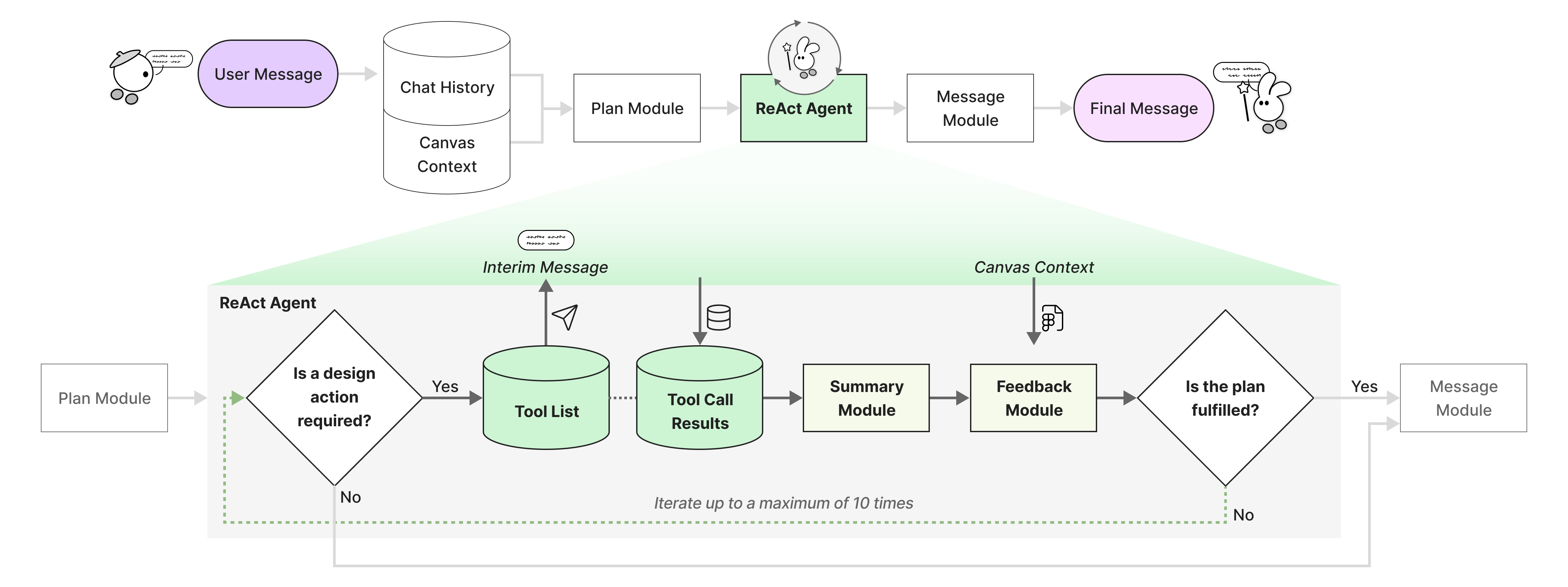}
    \caption{Agent structure of the first probe system. Upon receiving user input, the Plan Module generates an action plan. The ReAct Agent then iterates through a cycle of reasoning with tool selection, execution, summary, and feedback evaluation until the plan is fulfilled or the maximum of 10 iterations is reached. The Message Module generates the final response, reflecting both the agent's actions and the user's original message, and delivers it to the user.}
    \Description{Two levels of a flow diagram show the first probe's architecture. At the top, the User Message enters Chat History and Canvas Context, followed by the Plan Module, ReAct Agent, Message Module, and Final Message. The expanded ReAct loop below first asks whether a design action is required. If not, execution proceeds directly to the Message Module. If an action is required, the agent selects from the Tool List, produces an interim message, and obtains Tool Call Results. These results pass through the Summary Module and then the Feedback Module, which also receives Canvas Context. A second decision asks whether the plan is fulfilled. Yes leads to the Message Module; No follows a dashed return arrow to the first decision, with a maximum of ten iterations. The Message Module generates the final response reflecting the agent's actions and the original user request.}
    \label{fig:system1pipe}
\end{figure}

\clearpage
\section{Study 2 Interview Protocols}
\label{appendix:study2_interview_protocols}

\subsection{Stimulated Recall Interview}
Participants reviewed their visualized interaction logs and corresponding video clips. The following questions were asked for each segment.
\begin{enumerate}
    \item What are you doing in this video clip?
    \item What is the reason or purpose behind this behavior?
    \item In what collaborative situations do you think this behavior would be needed?
\end{enumerate}

\subsection{Wrap-up Interview (20 Minutes)}

\paragraph{Overall Collaboration Experience}
\begin{enumerate}
    \item How would you describe your overall experience of collaborating with the agent?
    \item Did your way of working with the agent change between Task 1 and Task 2? If so, how and why?
    \item How did working with this agent compare with your experience using other design agents?
\end{enumerate}

\clearpage
\section{Supplementary Results}
All results reported in this section are derived from Study 2.
\subsection{Interaction Log Processing and Visualization}
\label{appendix:vis_method}
\begin{figure}[h]
  \centering
  \includegraphics[width=\textwidth]{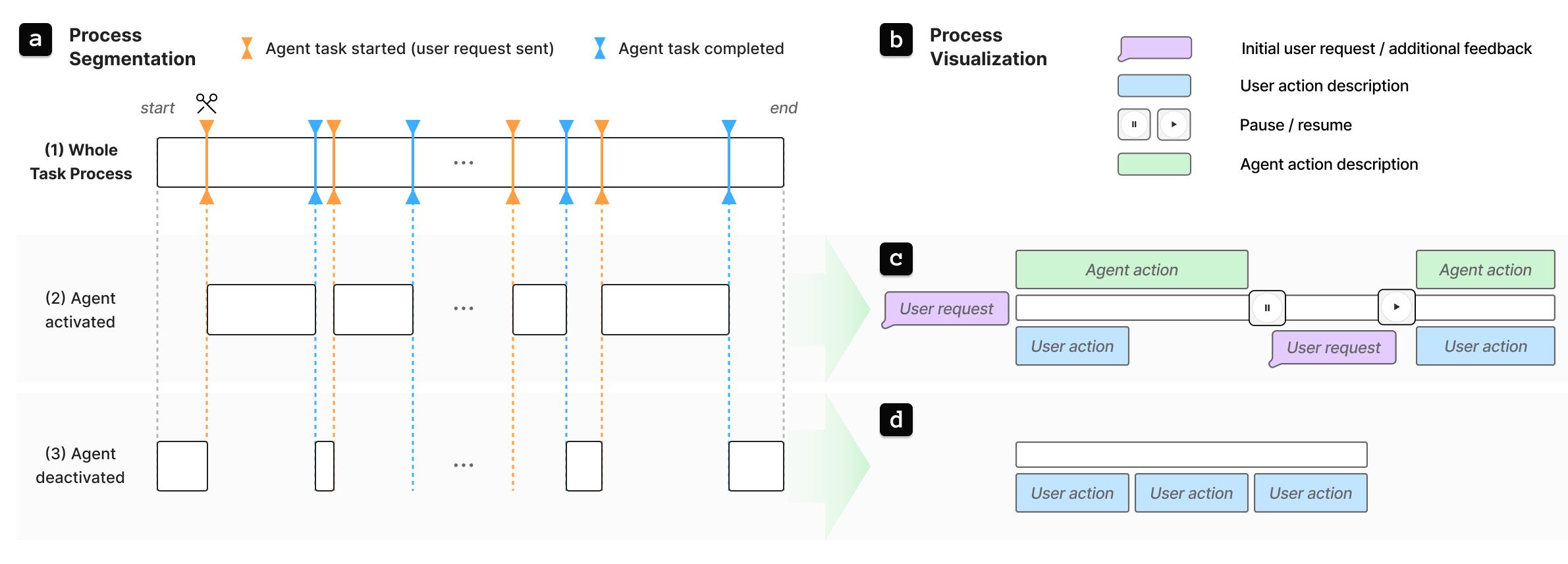}
  \caption{Visualization of interaction logs. (a) The overall task process is segmented into periods when the agent is activated and deactivated, using the timestamps of agent task initiation (triggered by user requests) and task completion. (b) Each segment is visualized with labeled interaction events. (c) Example of a visualized segment during agent activation. (d) Example of a visualized segment during agent deactivation.}
  \Description{Four panels show how interaction logs are segmented and visualized. In (a), orange markers identify agent task starts triggered by user requests, and blue markers identify task completions along the whole task timeline. Dashed guides separate the timeline into agent-activated intervals between these markers and the remaining agent-deactivated intervals. In (b), the legend maps purple speech bubbles to initial requests or additional feedback, blue rectangles to user action descriptions, green rectangles to agent action descriptions, and pause and play icons to pause and resume events. In (c), an activated segment begins with a user request. Agent action blocks above the timeline overlap user action blocks below it. A pause icon, an additional request, and a resume icon precede further agent and user actions. In (d), a deactivated segment contains three successive user action blocks and no agent action blocks.}
  \label{fig:vis_method}
\end{figure}

\clearpage
\subsection{User Action Pattern Visualization}
\label{appendix:action_pattern}
\begin{figure}[!h]
  \centering
  \includegraphics[width=0.9\textwidth]{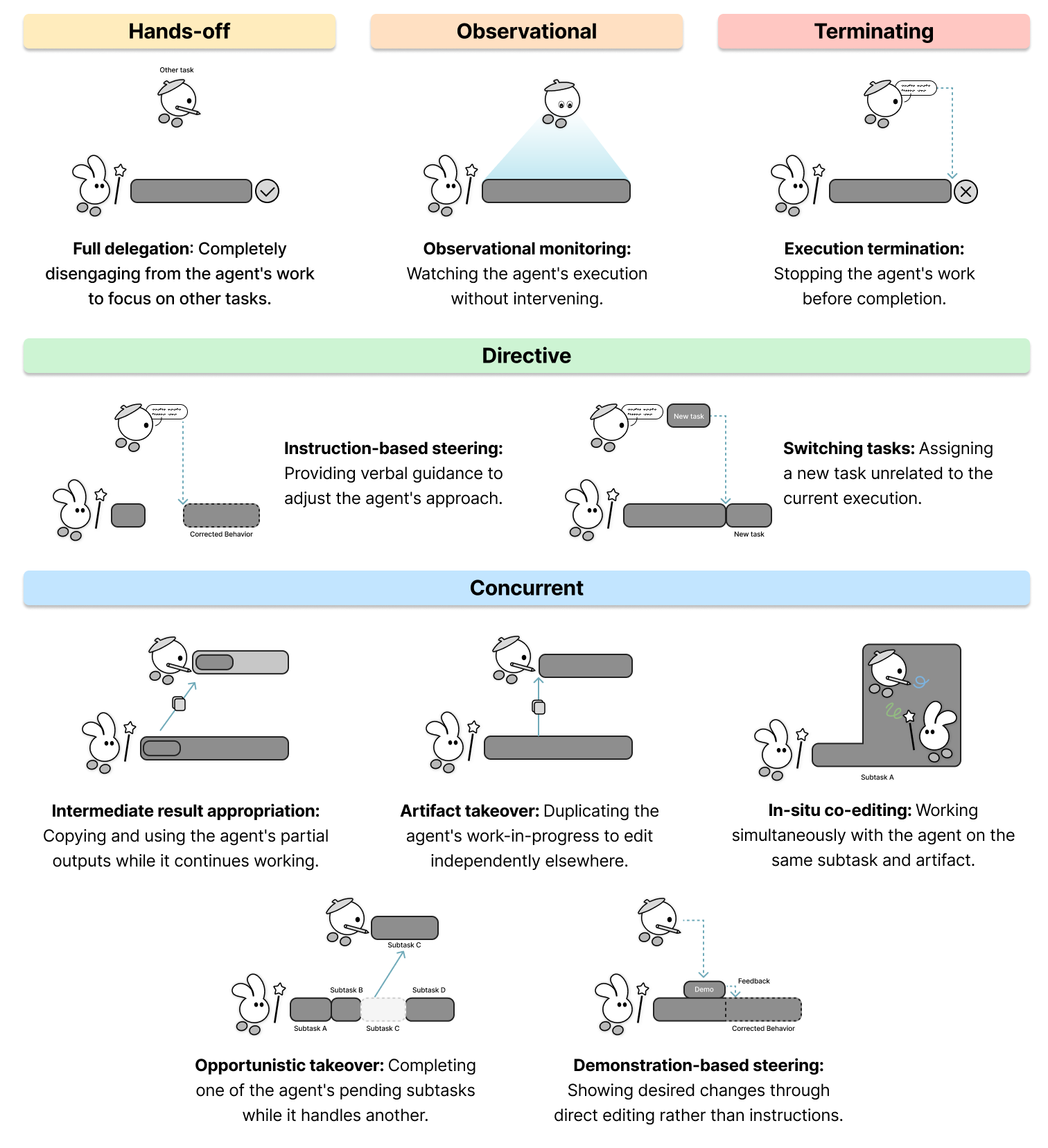}
  \caption{Visual representation of user action patterns.}
  \label{fig:action_pattern}
  \Description{Ten schematic action patterns are grouped into five categories. Beret-wearing characters represent users, rabbits represent agents, and gray shapes represent work or subtasks. Hands-off shows Full delegation: the user attends to another task while the agent works. Observational shows Observational monitoring: the user watches without intervening. Terminating shows Execution termination: the user stops the agent before completion. Directive contains Instruction-based steering, where verbal guidance changes the agent's approach, and Switching tasks, where the user assigns an unrelated new task. Concurrent contains five distinct patterns. Intermediate result appropriation shows the user copying part of the agent's output into their own work while the agent continues. Artifact takeover shows a separate copy of the agent's work-in-progress that the user edits independently. In-situ co-editing places user and agent together on the same subtask and artifact. Opportunistic takeover shows the user completing pending Subtask C while the agent handles other subtasks. Demonstration-based steering shows a direct edit labeled Demo feeding into the agent's subsequent corrected behavior.}
\end{figure}

\clearpage
\subsection{Turn-Level Action Category Distributions}
\label{appendix:distribution}
\begin{table}[h]
\resizebox{0.35\columnwidth}{!}{%
\begin{tabular}{lrl}
\hline
\rowcolor[HTML]{EFEFEF} 
\cellcolor[HTML]{EFEFEF}\textbf{Combination} & \multicolumn{1}{l}{\cellcolor[HTML]{EFEFEF}\textbf{Count}} & \textbf{\%} \\ \hline
(H) & 76 & 29.80\% \\
(O) & 36 & 14.12\% \\ \hline
(O) + (H) & 28 & 10.98\% \\
(O) + (C) & 14 & 5.49\% \\
(O) + (D) & 7 & 2.75\% \\
(O) + (T) & 3 & 1.18\% \\ \hline
(O) + (C) + (H) & 22 & 8.63\% \\
(O) + (D) + (H) & 14 & 5.49\% \\
(O) + (C) + (D) & 14 & 5.49\% \\
(O) + (C) + (T) & 2 & 0.78\% \\
(O) + (H) + (T) & 3 & 1.18\% \\ \hline
(O) + (C) + (H) + (T) & 3 & 1.18\% \\
(O) + (C) + (D) + (H) & 26 & 10.20\% \\
(O) + (D) + (H) + (T) & 2 & 0.78\% \\
(O) + (C) + (D) + (T) & 2 & 0.78\% \\ \hline
(O) + (C) + (D) + (H) + (T) & 3 & 1.18\% \\
\end{tabular}%
}
\captionsetup{justification=centering}
\caption{Turn-level distribution of action category combinations across 255 turns.}
\Description{The table lists 16 combinations of action categories across 255 turns, with a count and percentage for each combination. H means Hands-off, O Observational, C Concurrent, D Directive, and T Terminating. A combination identifies the categories present within a turn, not their order. Hands-off only is the most frequent combination; every other listed combination includes Observational. The rows are: Hands-off only, 76 turns (29.80\%); Observational only, 36 (14.12\%); Observational and Hands-off, 28 (10.98\%); Observational and Concurrent, 14 (5.49\%); Observational and Directive, 7 (2.75\%); Observational and Terminating, 3 (1.18\%); Observational, Concurrent, and Hands-off, 22 (8.63\%); Observational, Directive, and Hands-off, 14 (5.49\%); Observational, Concurrent, and Directive, 14 (5.49\%); Observational, Concurrent, and Terminating, 2 (0.78\%); Observational, Hands-off, and Terminating, 3 (1.18\%); Observational, Concurrent, Hands-off, and Terminating, 3 (1.18\%); Observational, Concurrent, Directive, and Hands-off, 26 (10.20\%); Observational, Directive, Hands-off, and Terminating, 2 (0.78\%); Observational, Concurrent, Directive, and Terminating, 2 (0.78\%); and all five categories, 3 (1.18\%).}
\label{tab:distribution}
\end{table}

\end{document}